\documentclass[aps,superscriptaddress,eqsecnum,nofootinbib,showpacs,
preprintnumbers]{revtex4}

\usepackage{amsmath}
\usepackage{amsfonts}
\usepackage{amssymb}
\usepackage{mathtools}
\usepackage{bm}
\usepackage{xcolor}
\usepackage{hyperref}
\usepackage{orcidlink}

\usepackage{graphicx}
\usepackage{subcaption}

\usepackage{color}
\usepackage{relsize}

\newcommand{\be}{\begin{equation}}
\newcommand{\ee}{\end{equation}}
\newcommand{\bea}{\begin{eqnarray}}
\newcommand{\eea}{\end{eqnarray}}

\begin{document}

\title{Generalizing the interacting dilatonic ghost condensate as a dark energy model.}

\author{Manuel Gonzalez-Espinoza\orcidlink{0000-0003-0961-8029}}
\email{manuel.gonzalez@upla.cl}
\affiliation{Laboratorio de investigación de Cómputo de Física, Facultad de Ciencias Naturales y
Exactas, Universidad de Playa Ancha, Subida Leopoldo Carvallo 270, Valparaíso, Chile}

\author{Ram\'on Herrera\orcidlink{0000-0002-6841-1629}}
\email{ramon.herrera@pucv.cl}
\affiliation{Instituto de F\'{\i}sica, Pontificia Universidad Cat\'olica de 
Valpara\'{\i}so, 
Casilla 4950, Valpara\'{\i}so, Chile}

\author{Johan Casimiro\orcidlink{0009-0008-4759-5685}}
\email{j.casimiro@pucp.edu.pe}
\affiliation{Pontificia Universidad Cat\'olica del Per\'u, Av.\ Universitaria 1801, San Miguel, Per\'u.}

\date{\today}

\begin{abstract} 
In this article, we study the cosmic evolution of a generalized dilatonic ghost condensate field as a dark energy candidate, formulated from a Lagrangian density with two dominant kinetic terms; one linear and one of arbitrary integer $n>2$ in combination with an exponential potential,
which interacts with dark matter through a source term. We analyzed three scenarios: the non-interacting situation $Q=0$ and two different interaction models, $Q\propto\rho_m\dot{\phi}$ and $Q\propto \rho_m H$ to describe the evolution of the present universe. For each interaction $Q$, we perform a detailed phase-space analysis to obtain stability conditions and identify critical points. In all situations, the system reproduces the standard cosmological dynamics and evolves toward late-time dark energy-dominated attractors, with quintessence or phantom features depending on the sign of the coupling parameter $\alpha$ associated with the standard kinetic term.
Furthermore, a joint likelihood analysis with Cosmic Chronometers, PantheonPlus, and DESI observations is performed for two values of power $n$ ($n=3$ and $n=5$) to determine marginalized parameter constraints at the confidence levels of 68$\%$ and 95$\%$ for the different $Q-$models.
For the interaction term $Q\propto \dot{\phi}\rho_m$, we find that the direction of the flow of energy depends on the sign of the coupling parameter $\alpha$ associated with the standard kinetic term. However, for the interaction $Q\propto H\,\rho_m$, the direction of the energy flow is independent of the sign of the coupling parameter $\alpha$ and always remains negative, corresponding to an energy transfer from dark matter to dark energy.
 
\end{abstract}


\maketitle

\section{Introduction}\label{Introduction}

In modern cosmology, the origin of the current accelerated expansion of the universe remains one of the most fundamental open questions. Based on observations of type-Ia supernovae \cite{Riess1998,Perlmutter1999}, this acceleration has been described in cosmology as the consequence of a dominant energy component with negative pressure, called dark energy (DE). The simplest energy density is the constant $\Lambda$, which combined with cold dark matter (CDM), gives rise to the $\Lambda$CDM model \cite{Weinberg1989,Peebles2003}.
However, the $\Lambda$CDM model presents some well-known theoretical difficulties \cite{Copeland2006}, which have motivated different explorations related to the dynamical dark energy model.

To describe the energy density associated with dark energy, the most studied alternatives are scalar fields. In particular, the quintessence stage established on the canonical kinetic term produces a simple consistent framework in which the equation of state (EoS) related to the DE, $w_{de}$, evolves dynamically to values of $w_{de}\simeq -1$ \cite{Ratra1988,Chiba2000}. In this context, there are more general models that arise when non-standard kinetic terms are added to the Lagrangian density related to the scalar field, leading to the broad class of k-essence models\cite{ArmendarizPicon1999,Garriga1999}, which can generate different cosmological behaviors in the dynamics of the  DE density.  Regarding  the theoretical framework, Lagrangian densities with multiple kinetic terms emerge naturally in low-energy effective actions of string theory, in which loop corrections generate a non-trivial dependence of the kinetic-term coefficient on moduli fields, see e.g. Refs.\cite{Damour1994,Damour1996,Foffa1999}.

In relation to the interaction between DE and DM, it has received attention in the literature, originally as a possible mechanism to alleviate the coincidence problem \cite{Amendola:2006dg,delCampo2006,delCampo2008,delCampo2009,Herrera2016,Amendola2000,
Chimento2003,Wang2016,He2008,Yang2018,Herrera2004,delCampo2005a,delCampo2005b}.   However, from a theoretical framework, interacting DE-DM models are well motivated in the context of quantum field theory in curved spacetime, where couplings between scalar fields and matter may emerge through radiative corrections or arise in effective descriptions of infrared-modified theories  \cite{Linde:1982zj, Birrell:1982ix}. From a phenomenological viewpoint, introducing an interaction between the dark sectors modifies the standard energy-momentum conservation equations from the source term $Q$, which accounts for the exchange between DE and DM \cite{Amendola:1999qq}. This framework  has been extensively investigated and gives rise to a wide variety of cosmological behaviors, including scaling solutions, attractors, and scenarios where the equation-of-state parameter crosses the phantom divide  without introducing  exotic degrees of freedom or fields \cite{He:2008si, Valiviita:2008iv}.
Thus, in the literature, different forms of the coupling term $Q$ have been proposed and analyzed. In this sense, various coupling term $Q$ can significantly affect the dynamics and the stability properties of the dynamical system (critical point).

If the interaction term $Q$ (see Eqs. (\ref{ecde}) and (\ref{rho_m})) is considered positive, the energy is transferred from DE to DM and  both components evolve differently from the standard cosmological stage. In this context, the DE contribution becomes more significant at earlier epochs, which may alleviate the cosmic coincidence problem. This direction of the flow modifies the expansion evolution and generally favors lower values of the Hubble rate at late times, which can intensify the current  Hubble tension. In addition, this flow of energy suppresses the growth of matter perturbations, alleviating the $S_8 $ discrepancy \cite{Lucca:2021dxo}. Furthermore, this direction of energy flow is consistent with the laws of thermodynamics \cite{Pavon:2007gt}.  When energy is transferred from DM to DE ($Q<0$), the cosmological evolution is modified significantly from the case of energy flow in the opposite direction. These interacting dark-sector models have attracted considerable attention because they can reduce the current Hubble tension, see e.g. Refs.\cite{Valiviita:2009nu,Kumar:2016zpg,Murgia:2016ccp,DiValentino:2017iww,Pan:2023mie,Yang:2025uyv,Dai:2026pvx,Figueruelo:2026eis}.

In relation to the theoretical framework beyond canonical theory, 
the dilatonic ghost condensate is a particular model associated with the k-essence, initially proposed as a consistent realization of phantom-like dynamics without ghost instability \cite{Piazza2004}.  In this model, the Lagrangian density includes both a linear kinetic term and a quadratic kinetic term with a dilatonic coupling, and the resulting dynamics allow for the EoS parameter $w_{de}<-1$, without compromising stability conditions. The analysis of the dynamics of this model has been made in detail through a dynamical system, whose properties  ensure the cosmological evolution  \cite{Copeland2006}.

An analysis of the interacting dilatonic ghost condensate, where the scalar field exchanges energy with DM from a coupling interaction term $Q$, was developed recently in Ref.\cite{Gonzalez-Espinoza:2025bzd}. In this study, three different interaction models were explored to describe the present universe. For each interaction model, a detailed phase-space analysis was developed, obtaining the stability conditions and identifying the critical points.
In addition, the different interaction models were compared with the most recent Hubble parameter and supernova Ia data as functions of redshift. Besides, the conditions for scaling regimes in these interacting models were analyzed together with the successful transition toward an attractor point to characterize the behavior of dark matter.

Although both dilatonic ghost condensate scenarios and interacting dark energy models have been extensively studied independently, their combined cosmological implications have been investigated for a particular case recently in Ref.\cite{Gonzalez-Espinoza:2025bzd}. The modification of the Lagrangian related to the scalar field naturally gives rise to scalar field dynamics beyond the canonical theory, leading to a rich phenomenology at both dynamical levels (background and perturbative). At the same time, interactions between DM and DE generate the possibility of modifying the late-time cosmic dynamics, alleviating some of the  current observational disagreement. In this sense, studying a cosmological scenario that includes both components may reveal novel dynamical stages and distinctive effects  that are missing  in the canonical-interaction model.

In this paper, we study the cosmological implications of a generalized dilatonic ghost condensate scalar field as a candidate for DE, while interacting with DM through a phenomenological coupling term $Q$. In this context, we analyze how different interactions that enter the energy balance equations for the non-canonical scalar field associated with DE and DM affect the evolution of the late-time evolution of the universe. In this sense, we consider  a non-interaction case and  two situations of the interaction kernel $Q$ that have been proposed in the literature. For each case,  we analyze the dynamical system and then identify critical points and their stability, allowing us to classify the cosmic evolution related with each non-canonical interaction model. In addition, we perform a statistical analysis using  Hubble parameter measurements and Type Ia supernova data from the PantheonPlus sample,  Cosmic Chronometers and Baryon Acoustic Oscillations from the DESI-DR2 survey to assess its phenomenological viability and to constrain the free parameters of each non-canonical interaction model. 

The article is organized as follows. In Sec.\ref{model} we describe the theoretical model and derive the fundamental equations. Sec.\ref{Ds}, we describe the dynamical system in our interacting model, and we study the analysis of critical points and their stability for each model. The statistical analysis and observational constraints are shown in Sec.\ref{sec8}. Finally, in Sec.\ref{conclusion} we present our conclusion. In this work, we chose units such that $c=\hbar=1$.

\section{Interaction in a  generalized dilatonic ghost condensate theory}\label{model}

In order to describe our model, we begin with a general four-dimensional effective action  defined  as \cite{Armendariz-Picon:1999hyi,Garriga:1999vw}
\bea
S&=&\int \Bigg[\frac{R}{2\kappa^2} +\mathcal{L}(X,\phi)
\Bigg]\,\sqrt{-g}\,d^{4}{x} 
\, + S_{m}+ S_{r},\label{action}
\eea 
where $\mathcal{L}(X,\phi)$ denotes the Lagrangian density associated with the scalar field and its kinetic term, the kinetic term being defined as $X= -\frac{1}{2} \partial_{\mu}{\phi}\partial^{\mu}{\phi}$, where $\phi$ represents the scalar field. The quantities $S_m$ and $S_r$ correspond to the actions of matter and radiation, respectively. Furthermore, $R$ is the Ricci scalar, $g$ is the determinant of the metric, and in our system of units, the parameter $\kappa^2$ is defined as $\kappa^2 = 8\pi / M_p^2$, where $M_p$ denotes the Planck mass.

Based on the effective action given in Eq. (\ref{action}), the energy density $\rho_{de}$ and pressure $p_{de}$ associated with dark energy (DE), assuming that it behaves as a perfect fluid, can be expressed as functions of the scalar field and the kinetic term \cite{Armendariz-Picon:1999hyi, Garriga:1999vw} 
\begin{equation}
\rho_{de} = 2X \frac{\partial \mathcal{L}(X, \phi)}{\partial X} - \mathcal{L}(X, \phi), \quad \text{and} \quad p_{de} = \mathcal{L}(X, \phi). \label{rp}
\end{equation}
It is worth noting that, in the particular case where the Lagrangian $\mathcal{L}(X, \phi) = X - V(\phi)$, with $V(\phi)$ representing the effective potential of the scalar field, these expressions reduce to standard results for the energy density and pressure of a scalar field within the framework of canonical theory \cite{Chiba:1999ka, Ratra:1987rm}.

In scalar field theories, the Lagrangian density $\mathcal{L}(X,\phi)$ can be generalized to include non-standard kinetic terms. A common approach is to express it as a series expansion in powers of the kinetic term $X$, such that \cite{Armendariz-Picon:1999hyi}
\begin{equation}
\mathcal{L}(X,\phi) = \sum_{n \geq 1} \alpha_n(\phi) X^{n} - V(\phi),
\label{exp1}
\end{equation}
where the coefficients $\alpha_n(\phi)$ are arbitrary functions of the scalar field $\phi$, the exponent $n$ is a positive integer, and the sum encodes the dynamics of the model beyond the canonical case. 
For a fixed value of the kinetic term $X$, the dominant contribution to the series is expected to arise from a specific term, depending on the behavior of the functions $\alpha_n(\phi)$. Consequently, it is reasonable to assume that, within a certain range of $X$, the Lagrangian can be effectively approximated by two dominant terms
\begin{equation}
\mathcal{L}(X,\phi) = \alpha X+\alpha_n(\phi) X^{n} - V(\phi),\,\,\,\,\,\mbox{with}\,\,\,\,\,n\geq 2.
\label{exp2}
\end{equation}
Here we consider that the parameter $\alpha_1=\alpha$ corresponds to a dimensionless constant positive or negative,   and that the coupling function $\alpha_n(\phi)$ has units of $M_p^{4(1-n)}$. In the standard case of canonical theory, we set
$\alpha=1$ and the coupling function
$\alpha_n(\phi)=\alpha_n=0$, so that the Lagrangian reduces to the usual form with a linear kinetic term. In the following, for convenience, we will use the notation for the coupling function $\alpha_n(\phi)$ as  $\alpha_n(\phi)=\alpha_n$.

The Lagrangian density in Eq.(\ref{exp2}) is motivated by a generalization of the dilatonic ghost condensate theory, which corresponds to the specific case $n=2$ together with the coupling parameter $\alpha_1=\alpha=-1$ \cite{Copeland:2006wr,Arkani-Hamed:2003juy,Gonzalez-Espinoza:2025bzd}. Additionally, this Lagrangian may arise from loop corrections in string theory, which introduce a non-trivial dependence of the kinetic term coefficients on the moduli fields, see e.g. Refs.\cite{Damour:1994zq,Damour:1995pd,Foffa:1999dv}.

Thus, inspired by these low-energy effective actions, we consider a minimal model to explore the interaction between dark energy, represented by a scalar field whose energy density is defined by Eq.~(\ref{rp}), with a Lagrangian density given by Eq.~(\ref{exp2}), and energy dark matter, characterized by its energy density, denoted by $\rho_m$.

To analyze the cosmological background dynamics of our model, we assume that the scalar field is homogeneous, that is, $\phi = \phi(t)$, and then the kinetic term associated with the scalar field becomes $X=\dot{\phi}^2/2$. We also consider a spatially flat Friedmann-Lemaître-Robertson-Walker (FLRW) metric, which can be written as
\begin{equation}
ds^2 = -dt^2 + a^2(t)\, \delta_{ij} dx^i dx^j ,
\label{FRWMetric}
\end{equation}
where
$a(t)$ is the scale factor that depends on cosmic time
$t$.

In this form, from Eqs. (\ref{rp}) and (\ref{exp2}), we find that the effective energy density $\rho_{de}$ and pressure $p_{de}$ associated with dark energy as a function of the scalar field are given by
\bea
\label{rhode}
\rho_{de}&=\alpha\,X+ (2n-1)\alpha_n\,X^n+V=\alpha \frac{\dot{\phi}^2}{2}+(2n-1)\alpha_n\,\left(\frac{\dot{\phi}^2}{2}\right)^n+V,
\eea
and
\bea
 p_{de}=\mathcal{L}(X,\phi)=\alpha\,X+\alpha_n\,X^n-V=\alpha \frac{\dot{\phi}^2}{2}+\alpha_n\,\left(\frac{\dot{\phi}^2}{2}\right)^n-V,\label{fm}
 \eea 
respectively.

Now, the Friedmann equation can be written in terms of the individual energy densities as
\bea
\label{SH00}
&& \frac{3}{\kappa^2} H^2=\rho_{tot}=\rho_{de}+\rho_{m}+\rho_{r},
\eea
and   we can also write the equation for $\dot{H}$ in terms of the densities and the pressure $p_{de}$  as
\bea
 -\frac{2}{\kappa^2} \dot{H}=\rho_{de}+p_{de}+\rho_{m}+\frac{4}{3}\rho_{r}.
\label{SHii}
\eea 
Here, $H=\dot{a}/a$ denotes the Hubble parameter, and $\rho_{tot}$ corresponds to the total energy density. Also, the densities $\rho_m$ and $\rho_r$ are the energy densities associated with matter, as mentioned before,  and radiation, respectively. 
In the following, the dots denote differentiation with respect to cosmological time $t$.

Furthermore, we can express the equation associated with the conservation of the total energy density as
\bea
\dot{\rho}_{tot}+3H(\rho_{tot}+p_{tot})=0.
\eea
In this framework, we may consider the existence of an interaction between DE and DM. This interaction can be modeled by introducing a coupling through a source term
$Q$, which enters the corresponding energy conservation equations for each component. In this sense, the equations of motion considering an interaction between the DE and DM can be written as
\begin{eqnarray}
\dot{\rho}_{de}+3H(\rho_{de}+p_{de})=-Q,\label{ecde}
\end{eqnarray} 
and
\bea
\label{rho_m}
 \dot{\rho}_{m}+3 H\rho_{m}=Q.
\eea

As mentioned before, the sign of the term $Q$ associated with the interaction determines the direction of energy exchange within the dark sector.   If $Q>0$, it indicates that dark energy acts as a source for dark matter, whereas if the interaction term $Q<0$, it implies that dark matter transfers energy to the dark energy component. In this sense, the direction of this flow of energy influences the cosmological evolution, modifying both the background dynamics and the perturbative evolution.

Additionally, for the radiation field, the equation of motion becomes 
\bea
\dot{\rho}_{r}+4 H\rho_{r}=0,
\label{rho_r}
\eea
and the solution for the radiation energy density $\rho_r$ is given by $\rho_r\propto a^{-4}$.

From Eqs. (\ref{fm}) and (\ref{ecde}), we can write the equation of motion for the scalar field as 

\bea
\left[\alpha+n(2n-1)\alpha_n\,\left(\frac{\dot{\phi}^2}{2}\right)^{n-1}\right]\ddot{\phi}+3H\dot{\phi}\left[\alpha+n\,\alpha_n\,\left(\frac{\dot{\phi}^2}{2}\right)^{n-1}\right]+(2n-1)(\alpha_n)_{,\phi} 
\left(\frac{\dot{\phi}^2}{2}\right)^{n}
+V_{,\phi} =-\dfrac{Q}{\dot{\phi}}.
\label{MFreq2}
\eea 
Here, the notation $V_{,\phi}$ corresponds to $\partial V/\partial \phi$ , $\alpha_n(\phi)=\alpha_n$ and $(\alpha_n(\phi))_{,\phi}=(\alpha_n)_{,\phi}=\partial \alpha_n/\partial \phi$.

In addition, from Eq.(\ref{fm}) we can obtain the EoS parameter $w_{de}$, related to the dark energy  as 
\begin{equation}
w_{de}=\frac{p_{de}}{\rho _{de}}=\frac{\alpha\,\frac{\dot{\phi^2}}{2}+\alpha_n\,\left(\frac{\dot{\phi}^2}{2}\right)^n-V}{\alpha\,\frac{\dot{\phi^2}}{2}+(2n-1)\alpha_n\,\left(\frac{\dot{\phi}^2}{2}\right)^n+V}.
\label{wDE1}
\end{equation}
Moreover, it is useful to define the total equation-of-state parameter $w_{\text{tot}}$ as
\be
w_{tot}=\frac{p_{tot}}{\rho_{tot}}=\frac{p_{de}+p_r}{\rho _{de}+\rho _m+\rho _r}=\frac{\alpha\,\frac{\dot{\phi^2}}{2}+\alpha_n\,\left(\frac{\dot{\phi}^2}{2}\right)^n-V+p_r}{\alpha\,\frac{\dot{\phi^2}}{2}+(2n-1)\alpha_n\,\left(\frac{\dot{\phi}^2}{2}\right)^n+V+\rho _m+\rho _r}.
\label{wtot}
\ee 
In terms of the total EoS parameter, one can establish a relation with the deceleration parameter $q$ by means of the following, 
$q=-\frac{\ddot{a}a}{\dot{a}^2}=\frac{1}
{2}\left(1+3w_{tot}\right).$
Accordingly, the condition for an accelerating universe $\ddot{a} > 0$, and then the deceleration parameter satisfies the condition $q < 0$, or, in terms of the total EoS parameter, when $w_{\text{tot}} < -1/3$.

Furthermore, we define the dimensionless density parameters corresponding to DM, DE, and radiation as
\bea
&& \Omega_{m}\equiv\frac{\kappa^2 \rho_{m}}{3 H^2},\:\:\:\: \Omega_{de}\equiv\frac{\kappa^2 \rho_{de}}{3 H^2},\:\:\:\: \Omega_{r}\equiv \frac{\kappa^2 \rho_{r}}{3 H^2}.
\eea 
Thus, we write  the  constraint imposed by the Friedmann equation (\ref{SH00}), which becomes
\be
\Omega_{de}+\Omega_{m}+\Omega_{r}=1.\label{ec}
\ee 
In what follows, we analyze the dynamical system corresponding to our model.

\section{Dynamical system}\label{Ds}

In this section, we investigate the dynamical system associated with our model, focusing on the identification of its critical points as well as the corresponding cosmological parameters and the stability properties of the autonomous system. To this end, we introduce the following dimensionless variables, defined as \cite{Copeland:2006wr,Hussain:2025uye,Chakraborty:2019swx,Dinda:2023mad}

\begin{equation}
\begin{aligned}
x =& \dfrac{\kappa  \dot{\phi}}{\sqrt{6} H}, \ \ \ \ \ \ &y =& \dfrac{\kappa  \sqrt{V}}{\sqrt{3} H}, \ \ \ \ \ \ \ \varrho=\frac{\kappa\sqrt{\rho_r}}{\sqrt{3}H},\\ 
u =& \dfrac{\kappa \sqrt{\alpha_n}}{\sqrt{3 } H} \left(\dfrac{\dot{\phi}}{\sqrt{2}}\right)^n, \ \ \ \ \ \
&\lambda =& - \dfrac{V_{,\phi}}{\kappa V}, \ \ \ \ \ \ \ \ \ \ \ \ \ \ \\
\sigma =&  \dfrac{(\alpha_n)_{,\phi}}{\kappa \alpha_n},  \ \ \ \ \ \ \ \ \ \ \ \ &\Gamma =&  \dfrac{V V_{,\phi\phi}}{(V_{,\phi})^2}, \ \ \ \ \ \ \ \Theta = \dfrac{\alpha_n\, (\alpha_n)_{,\phi\phi}}{[(\alpha_n)_{,\phi}]^2}. 
\end{aligned} \label{vardyn}
\end{equation}

Hence, the dynamical system can be written in a general form using these dimensionless variables as
\begin{eqnarray}
\dfrac{d x}{d N} &=& f_1(x,y,u, \varrho),
\label{dinsyseq1}\\
\dfrac{d y}{d N} &=& f_2(x,y,u, \varrho),
\label{dinsyseq2}\\
\dfrac{d u}{d N} &=& f_3(x,y, u,\varrho),
\label{dinsyseq4}\\
\dfrac{d \varrho}{d N} &=& f_4(x,y,u, \varrho),
\label{dinsyseq5}\\
\dfrac{d \lambda}{d N} &=& -\sqrt{6} \left(\Gamma -1\right) \lambda ^2 x, \label{dinsyseq6}\\
\dfrac{d \sigma}{d N} &=& \sqrt{6} \left(\Theta -1\right) \sigma^2 x. \label{dinsyseq8}
\end{eqnarray}

Here, the functions $f_i(x,y,u,\varrho)$ with $i=1,2,3,4$ are determined by the specific interaction under consideration. Furthermore, we assume that the interaction term $Q$ can be written in terms of the dimensionless variables introduced above. Moreover,  the variable $N$ denotes the number of $e-$folds, which is defined by the scale factor as $N = \ln a$.

In addition, considering the above set of space variables, we can write that the density parameter related to dark energy $\Omega_{de}$ is given by
\begin{equation}
    \Omega_{de}=\alpha  x^2+(2 n-1) u^2+y^2.
\end{equation}

Analogously, the parameter associated with the equation of state of dark energy $w_{de}$ becomes
\begin{equation}
    w_{de}=\frac{u^2+\alpha  x^2-y^2}{(2 n-1) u^2+\alpha  x^2+y^2},
\end{equation}
and the total equation of state $w_{tot}$ in terms of the new dynamical variables can be written as
\begin{equation}
    w_{tot}=u^2+\alpha  x^2-y^2+\frac{\varrho ^2}{3}.
\end{equation}

In what follows, we examine different forms of interaction derived from the source term $Q$, and provide a comprehensive dynamical analysis for each case.

From the left-hand side of Eqs.~(\ref{ecde}) and (\ref{rho_m}), it follows that the interaction term $Q$ should be proportional to the energy densities $\rho_m$ or $\rho_{de}$, or to their sums, etc., multiplied by a quantity with dimensions of inverse time \cite{delCampo:2008jx,delCampo:2008sr,delCampo:2006vv,Herrera:2016uci}.
In the literature, different combinations of these quantities have been proposed, leading to interaction terms of the general form $Q = Q(\kappa \rho_m \dot{\phi}, \kappa \rho_{de} \dot{\phi}, H \rho_m, H \rho_{de}, \ldots)$. In this context, we will first focus on the non-interacting case, and subsequently, we will analyze the two simplest and most widely studied interaction models, namely:
\bea
Q=\beta\kappa\rho_m\dot{\phi},\,\,\,\,\,\,\,\,\mbox{and}\,\,\,\,\,\,\,\,\,\,\,Q=3 \beta \rho_{m} H.
\eea
Here, the parameter $\beta$ is a constant that represents a dimensionless parameter. As discussed previously, various forms of the interaction term $Q$ have been explored in the literature. For alternative expressions of $Q$, see e.g., Refs.~\cite{Amendola:1999er,Chimento:2003iea,Wang:2016lxa,He:2008tn,Yang:2018xlt,Herrera:2004dh,delCampo:2015vha,delCampo:2005tr}.

Furthermore, in order to fully define the autonomous system described by Eqs. (\ref{dinsyseq1})–(\ref{dinsyseq8}), it is necessary to specify the effective potential $V(\phi)$ and the coupling function $\alpha_n(\phi)$ as functions of the scalar field. Thus, following Refs.\cite{Piazza:2004df,Gonzalez-Espinoza:2025bzd}, we consider a dark energy model featuring a  scalar field interacting with dark matter, where the effective potential $V(\phi)$ and the coupling function $\alpha_n(\phi)$ related to the field are defined  by
\bea
V(\phi)=V_0\,e^{-\lambda\,\phi}, \,\,\,\,\,\,\mbox{and}\,\,\,\,\,\,\alpha_n(\phi)=\alpha_{0_n}\,e^{\sigma\,\phi},\label{VF}
\eea
where the parameters $V_0$, $\lambda$, $\alpha_{0_n}$, and $\sigma$ correspond to  different constants. In this setup, $V_0$ has dimensions of $ M_p^4$, while the dimension of the parameter $\alpha_{0_n}$ depends on the power $n$ and is given by $ M_p^{4(1-n)}$. The parameters $\lambda$ and $\sigma$ have  dimensions of $ M_p^{-1}$.

\subsection{Non-interacting  $Q =0$. Critical points and Stability analysis.}\label{inter_0}

In this subsection, we analyze the dynamical system for the first case, where we consider a non-interacting model, i.e., $Q=0$, along with the effective potential and the coupling function expressed in terms of the scale factor as defined in Eq.~(\ref{VF}). In this sense, we obtain that the functions $f_i(x,y,u,\varrho)$
associated to the case $Q=0$ are defined as

\begin{eqnarray}
    f_1 &=& \frac{x}{4 n^2 u^2-2 n u^2+2 \alpha  x^2} \bigg( 
 \, 2 n^2 u^2 \left(3 u^2 + 3 \alpha x^2 - 3 y^2 + \varrho^2 + 3 \right)  - n u^2 \left(3 u^2 + 3 \alpha x^2 + 2 \sqrt{6} \sigma x - 3 y^2 + \varrho^2 + 9 \right) \nonumber \\
&& + x \left( \alpha x \left(3 u^2 - 3 y^2 + \varrho^2 - 3 \right) + \sqrt{6} \left( \sigma u^2 + \lambda y^2 \right) + 3 \alpha^2 x^3 \right)
\bigg),\label{f10} \\
f_2 &=& \frac{1}{2} y \left(3 u^2+3 \alpha  x^2-\sqrt{6} \lambda  x-3 y^2+\varrho ^2+3\right), \\
f_3 &=& \frac{u}{4 n^2 u^2-2 n u^2+2 \alpha  x^2} \bigg(
 \, 2 n^2 u^2 \left(3 u^2 - 3 y^2 + 3 \alpha x^2 + \varrho^2 \right)
- n \left(3 u^4 + x \left(6 \alpha x - \sqrt{6} \lambda y^2 \right) 
+ u^2 \left(3 - 3 y^2 + 3 \alpha x^2 + \varrho^2 \right) \right) \nonumber \\
&& + \alpha x^2 \left(3 + 3 u^2 - 3 y^2 + 3 \alpha x^2 + \varrho^2 
+ \sqrt{6} \sigma x \right)
\bigg), \\
f_4 &=& \frac{1}{2} \varrho  \left(3 u^2+3 \alpha  x^2-3 y^2+\varrho ^2-1\right).\label{f40}
\end{eqnarray}

\begin{table*}[ht]
 \centering
 \caption{Critical points for the autonomous system in the non-interacting model, $Q=0 $. With $A_c=\sigma ^2 \left(-24 \alpha  n^4+72 \alpha  n^3-72 \alpha  n^2+4 n^2 \sigma ^2+24 \alpha  n-4 n \sigma ^2+\sigma ^2\right)$}.
\begin{center}
\begin{tabular}{c c c c c c c c c}\hline\hline
Name &  $x_c$ & $y_c$ & $u_{c}$ & $\varrho_{c}$  \\\hline
$\ \ \ \ \ \ \ \ a_{R} \ \ \ \ \ \ \ \ $ & $0$ & $0$ & $0$  & $1$ \\
$\ \ \ \ \ \ \ \ b_{R} \ \ \ \ \ \ \ \ $ & $\frac{2 \sqrt{\frac{2}{3}} (n-1)}{\sigma }$ & $0$ & $\frac{2  \sqrt{\frac{2}{3}} \sqrt{\alpha } (n-1)}{\sqrt{\left((n-2) \sigma ^2\right)}}$  & $\frac{\sqrt{(n-2) \sigma ^2-8 \alpha  (n-1)^3}}{\sqrt{(n-2) \sigma ^2}}$ \\
$\ \ \ \ \ \ \ \ c_{R} \ \ \ \ \ \ \ \ $ & $\frac{2 \sqrt{\frac{2}{3}}}{\lambda }$ & $\frac{2 \sqrt{\alpha }}{\sqrt{3} \lambda }$ & $0$  & $\frac{\sqrt{\lambda ^2-4 \alpha }}{\lambda }$ \\
$\ \ \ \ \ \ \ \ d_{SM}^{\pm} \ \ \ \ \ \ \ \ $ & $\frac{\pm 1}{\sqrt{\alpha }}$ & $0$ & $0$  & $0$ \\
$\ \ \ \ \ \ \ \ e_{M} \ \ \ \ \ \ \ \ $ & $0$ & $0$ & $0$  & $0$ \\
$\ \ \ \ \ \ \ \ f_{M} \ \ \ \ \ \ \ \ $ & $\frac{\sqrt{\frac{3}{2}}}{\lambda }$ & $\frac{\sqrt{\frac{3}{2}} \sqrt{\alpha }}{\lambda }$ & $0$  & $0$ \\
$\ \ \ \ \ \ \ \ g_{M} \ \ \ \ \ \ \ \ $ & $\frac{\sqrt{\frac{3}{2}} (n-1)}{\sigma }$ & $0$ & $\frac{ \sqrt{\frac{3}{2}} \sqrt{-\alpha } (n-1)}{\sigma }$  & $0$ \\
$\ \ \ \ \ \ \ \ h \ \ \ \ \ \ \ \ $ & $\frac{\lambda }{\sqrt{6} \alpha }$ & $\sqrt{1-\frac{\lambda ^2}{6 \alpha }}$ & $0$  & $0$ \\
$\ \ \ \ \ \ \ \ i \ \ \ \ \ \ \ \ $ & $0$ & $0$ & $\frac{1}{\sqrt{2 n-1}}$  & $0$ \\
$\ \ \ \ \ \ \ \ j^{\pm} \ \ \ \ \ \ \ \ $ & $\frac{(2 n-1) \sigma ^2\pm \sqrt{A_c}}{2 \sqrt{6} \alpha  (n-1)^2 \sigma }$ & $0$ & $\frac{\sqrt{\frac{\mp\sqrt{A_c}+12 \alpha  (n-1)^3+(1-2 n) \sigma ^2}{\alpha  (n-1)^4}}}{2 \sqrt{3}}$  & $0$ \\

\\ \hline\hline
\end{tabular}
\end{center}
\label{table1}
\end{table*}
\begin{table}[ht]
 \centering
 \caption{Cosmological parameters for the critical points in Table \ref{table1}. With $A_c=\sigma ^2 \left(-24 \alpha  n^4+72 \alpha  n^3-72 \alpha  n^2+4 n^2 \sigma ^2+24 \alpha  n-4 n \sigma ^2+\sigma ^2\right)$.}
\begin{center}
\begin{tabular}{c c c c c c}\hline\hline
Name &   $\Omega_{de}$ & $\Omega_{m}$ & $\Omega_{r}$ & $\omega_{de}$ & $\omega_{tot}$ \\\hline
$a_{R}$ & $0$ & $0$ & $1$ & $\frac{1}{3}$ & $\frac{1}{3}$ \\
$b_{R}$ & $\frac{8 \alpha  (n-1)^3}{(n-2) \sigma ^2}$ & $0$ & $1-\frac{8 \alpha  (n-1)^3}{(n-2) \sigma ^2}$ & $\frac{1}{3}$ & $\frac{1}{3}$ \\
$c_{R}$ & $\frac{4 \alpha }{\lambda ^2}$ & $0$ & $1-\frac{4 \alpha }{\lambda ^2}$ & $\frac{1}{3}$ & $\frac{1}{3}$ \\
$d_{SM}^\pm$ & $1$ & $0$ & $0$ & $1$ & $1$ \\
$e_{M}$ & $0$ & $1$ & $0$ & $0$ & $0$ \\
$f_{M}$ & $\frac{3 \alpha }{\lambda ^2}$ & $1-\frac{3 \alpha }{\lambda ^2}$ & $0$ & $0$ & $0$ \\
$g_{M}$ & $-\frac{3 \alpha  (n-1)^3}{\sigma ^2}$ & $1+\frac{3 \alpha  (n-1)^3}{\sigma ^2}$ & $0$ & $0$ & $0$ \\
$h$ & $1$ & $0$ & $0$ & $-1+\frac{\lambda ^2}{3 \alpha }$ & $-1+\frac{\lambda ^2}{3 \alpha }$ \\
$i$ & $1$ & $0$ & $0$ & $\frac{1}{2 n-1}$ & $\frac{1}{2 n-1}$ \\
$j^{\pm}$ & $1$ & $0$ & $0$ & $-\frac{\mp \sqrt{A_c}+6 \alpha  (n-1)^3+(1-2 n) \sigma ^2}{6 \alpha  (n-1)^3}$ & $-\frac{\mp\sqrt{A_c}+6 \alpha  (n-1)^3+(1-2 n) \sigma ^2}{6 \alpha  (n-1)^3}$ \\
\\ \hline\hline
\end{tabular}
\end{center}
\label{table2c}
\end{table}

Thus, from the functions $f_i$, we determine the critical points for the non-interacting case, using the potential and the coupling function defined in Eq.(\ref{VF}). 
In this context, the critical points are determined by imposing the conditions $dx/dN = dy/dN = du/dN = d\varrho/dN = 0$ in Eqs.(\ref{dinsyseq1})–(\ref{dinsyseq8}), using the functions $f_i$ defined in Eqs. (\ref{f10})–(\ref{f40}), respectively.

In addition, from the definition of dynamical variables in Eq.(\ref{vardyn}), the physically admissible quantities associated with the critical points must satisfy the requirements $y_c\ge 0$, $u_c\ge 0$, and $\varrho_c\ge 0$. In the following, the notation with the subscript ``c'' is used to designate a critical point.  
In this context, the critical points of our system associated with the non-interacting model ($Q=0$) are shown in Table \ref{table1}. 
Moreover, the corresponding values of the cosmological parameters are summarized in Table \ref{table2c}. We should note that the case $n=2$ was analyzed in Ref.\cite{Gonzalez-Espinoza:2025bzd} and must be treated separately to avoid singularities at the critical points. In what follows, we focus on the cases where the associated power 
$n$ is greater than two.

In these tables, the symbol $a_R$ denotes the critical point associated with the radiation epoch. Here, we find that the parameter $\Omega_r$ does not depend on the power $n$ associated with the kinetic term, and that the total EoS is given by $w_{tot}=1/3$. 
 It is also observed that this critical point does not depend on the parameter $\lambda$ related to the effective potential and $\sigma$ related to the coupling term.

The critical point $b_R$ represents another point during the radiation-dominated era, which depends on the parameters $n$, $\alpha$ and $\sigma$ (associated with the coupling function), where the EoS parameters $w_{de}=w_{total}=1/3$. Here we also note that there is a contribution related to dark energy $\Omega_{de}$ and this contribution depends on parameters $n$, $\alpha$, and $\sigma$. Also, we note that this point $b_R$ only exits if the parameter $\alpha\ge0$ and $\alpha \leq \frac{(n-2) \sigma ^2}{8 (n-1)^3}$, where the range for the parameter $\alpha$ is given by $\frac{(n-2) \sigma ^2}{8 (n-1)^3}\ge \alpha\ge 0$. In addition, we have a third point during the radiation-dominated epoch $c_R$, which depends on the parameters $\alpha$ and $\lambda$. Also, here, this point only exists if the parameter $\alpha$ is a positive quantity or equal to zero.

The point $d_{SM}^{\pm}$ represents a stiff matter where the total EoS parameter $w_{tot}=1$ and $w_{de}=1$, and this point exists only for $\alpha>0$  (unstable point). Here, the critical point $x_c$ has two values;  $x_c=\pm 1/\sqrt{\alpha}$ and the energy density associated with dark energy scales as $\rho_{de}\propto a^{-6}$.
Moreover, the critical points $f_{M}$, $e_M$, and $g_M$ represent solutions dominated by matter, where the EoS parameters are associated with dust $w_{de}=w_{tot}=0$. In these cases, the critical points $f_M$ and $g_M$ depend on the parameter $\alpha$. For these points, only $e_M$ has a parameter $\Omega_m=1$. In addition, we note that the point $f_M$ exists only if the coupling parameter $\alpha$ is positive or equal to zero. Here, the density parameters $\Omega_m$ and $\Omega_{de}$ depend on the parameters $\alpha$ and $\lambda$. 
In contrast to point $f_M$, the point  $g_M$ exists only if the parameter $\alpha\le0$. In addition,  we observe that the density parameters $\Omega_m$ and $\Omega_{de}$ depend on the parameters $\alpha$, the power $n$ and $\sigma$.

 For the critical points $h$, $i$ and $j^\pm$ represent points associated with the dark energy epoch. For each of these points,  we find that  $w_{de}=w_{tot}$ and these points exhibit the density parameter related to the dark energy $\Omega_{de}=1$. In particular, for the point $h$, we note that for positive values of $\alpha$ it corresponds to a quintessence-like behavior, while $\alpha<0$, it corresponds to the phantom regime. Also, we note that the point $j^\pm$ presents two values associated with  the critical points $x_c$ and $u_c$, see Table \ref{table1}.

 In summary, the system admits different critical points for both values of $\alpha>0$ and $\alpha<0$, each one associated with distinct dynamical behaviors.

On the other hand, to investigate the stability of the critical points, we introduce small, time-dependent linear perturbations in the dimensionless variables of the dynamical system around each critical point. In this context, we can consider that $x=x_c+\delta x$, $y=y_c+\delta y$, $\varrho=\varrho_c+\delta\varrho$, and  $u=u_c+\delta u$,  where the variables $\delta x$, $\delta y$,  $\delta \varrho$, and  $\delta u$,  denote  small perturbations such that $x_c\gg\delta x$, $y_c\gg \delta y$, etc.

By substituting these perturbations into Eqs. (\ref{dinsyseq1})–(\ref{dinsyseq8}), we construct the linear perturbation matrix $\mathcal{M}$, Ref. \cite{Copeland:2006wr}.
Using this approach, we compute the eigenvalues of the matrix $\mathcal{M}$, which, when evaluated at each fixed point, are labeled $\mu_i$ with $i=1,2,3,4$. Thus, we can determine the stability conditions of the various points by analyzing the eigenvalues $\mu_i$.

In the classification of critical point stability, the system corresponds to a stable node when all the eigenvalues are negative, whereas it corresponds to an unstable node when all the eigenvalues are positive. Additionally, within this classification, a saddle point arises when the eigenvalues have different signs, whereas a stable spiral occurs when the eigenvalues are complex with negative real parts.
In this context, it is worth noting that critical points identified as stable nodes or stable spirals act as attractors of the dynamical system \cite{Copeland:2006wr}.

In the following, we provide the eigenvalues and corresponding stability conditions of the critical points associated with the non-interacting model.
\\
\\
\\
\begin{itemize}
    \item Point $a_R$ has the eigenvalues   
    \begin{eqnarray}
        \mu_1 &=& -1, \ \ \ \ \ \mu_2 = 1, \ \ \ \ \ 
        \ \mu_3 = 2, \ \ \ \ \ \mu_4 = 2-3n ,\nonumber\\
        &&
    \end{eqnarray}
    therefore, it is a saddle point.
    \item Point $b_R$ exits for 
    \begin{equation}
        n > 2\land \alpha \leq \frac{(n-2) \sigma ^2}{8 (n-1)^3},
    \end{equation}
    and it has the eigenvalues   
    $$
        \mu_1 = 1, \ \ \ \ \ \mu_2 = \frac{2 (\lambda +\lambda  (-n)+\sigma )}{\sigma }, \ \ \ \ \ \nonumber\\
        \ \mu_3 = -\frac{\sqrt{(n+1) \sigma ^2 \left(3 (3 n-5) \sigma ^2-64 \alpha  (n-1)^3\right)}}{2 (n+1) \sigma ^2}-\frac{1}{2},
        $$
        and
      \begin{eqnarray}  
        \mu_4 = \frac{1}{2} \left(\frac{\sqrt{(n+1) \sigma ^2 \left(3 (3 n-5) \sigma ^2-64 \alpha  (n-1)^3\right)}}{(n+1) \sigma ^2}-1\right) ,\nonumber\\
        &&
    \end{eqnarray}

    therefore, it is unstable.
    \item Point $c_R$ exits for
    \begin{eqnarray}
       \left(\lambda <0\land 0\leq \alpha \leq \frac{\lambda ^2}{4}\right)\lor (\lambda =0\land \alpha =0)\lor \left(\lambda >0\land 0\leq \alpha \leq \frac{\lambda ^2}{4}\right)
    \end{eqnarray}
    it has the eigenvalues
    $$
        \mu_1 = 1, \ \ \ \ \ \mu_2 = \frac{1}{2} \left(-\frac{\sqrt{\alpha  \lambda ^2 \left(64 \alpha -15 \lambda ^2\right)}}{\sqrt{\alpha } \lambda ^2}-1\right), \ \ \ \ \ \mu_3 = \frac{1}{2} \left(\frac{\sqrt{\alpha  \lambda ^2 \left(64 \alpha -15 \lambda ^2\right)}}{\sqrt{\alpha } \lambda ^2}-1\right),  
        $$
        and
       \begin{equation} 
        \mu_4 = \frac{2 \sigma }{\lambda }-2 n+2,
    \end{equation}
    therefore, it is unstable.
    \item Point $d_{SM}^\pm$ exits for $\alpha >	0$ and it has the eigenvalues
    \begin{eqnarray}
        \mu_1 &=& 1, \ \ \ \ \ \mu_2 = 3, \ \ \ \ \ 
        \mu_3 = \frac{\sqrt{\frac{3}{2}} \lambda }{\sqrt{\alpha }}+3 \ \ \ \ \ 
        \mu_4 = -\frac{\sqrt{\frac{3}{2}} \sigma }{\sqrt{\alpha }}-3 n+3,
    \end{eqnarray}
    therefore, it is unstable.
    \item Point $e_M$ has the eigenvalues   
    \begin{eqnarray}
        \mu_1 &=& -\frac{1}{2}, \ \ \ \ \ \mu_2 = -\frac{3}{2}, \ \ \ \ \ 
        \ \mu_3 = \frac{3}{2}, \ \ \ \ \ \mu_4 = \frac{1}{2} (3-6 n) ,\nonumber\\
        &&
    \end{eqnarray}
    therefore, it is a saddle point.
    \item Point $f_{M}$ exits for $\alpha \geq 0$ and it has the eigenvalues
    \begin{eqnarray}
        \mu_1 &=& -\frac{1}{2},\ \ \ \ \ \ \
        \mu_2 = \frac{3 (\lambda +\lambda  (-n)+\sigma )}{2 \lambda }, \nonumber \\
        \mu_3 &=& \frac{3}{4} \left(-\frac{\sqrt{  \lambda ^8 \left(24 \alpha -7 \lambda ^2\right)}}{ \lambda ^5}-1\right), \ \ \ \ \ \ \ 
        \mu_4 = \frac{3}{4} \left(\frac{\sqrt{  \lambda ^8 \left(24 \alpha -7 \lambda ^2\right)}}{ \lambda ^5}-1\right),
    \end{eqnarray}

    therefore, it is a saddle point for
   \begin{align}
    &n\geq 2\land \alpha \geq 0\land 3 \alpha \leq \lambda ^2\land ((\lambda <0\land \lambda +\sigma <\lambda  n)\lor (\lambda >0\land \lambda +\sigma >\lambda  n)).\nonumber \\
    &
    \end{align}

    \item Point $g_{M}$ exits for $\alpha \leq 0$ and it has the eigenvalues
    \begin{eqnarray}
        \mu_1 &=& -\frac{1}{2},\ \ \ \ \ \ \
        \mu_2 = \frac{3 (\lambda +\lambda  (-n)+\sigma )}{2 \sigma }, \nonumber \\
        \mu_3 &=& -\frac{3 \left( \left(2 n^2-n-1\right) \sigma ^2+\sqrt{ \left(-(n-1)^2\right) (2 n+1) \sigma ^2 \left(24 \alpha  (n-1)^3+(7-2 n) \sigma ^2\right)}\right)}{4  (n-1) (2 n+1) \sigma ^2},\nonumber \\
        \mu_4 &=& \frac{3 \left( \left(-2 n^2+n+1\right) \sigma ^2+\sqrt{  \left(-(n-1)^2\right) (2 n+1) \sigma ^2 \left(24 \alpha  (n-1)^3+(7-2 n) \sigma ^2\right)}\right)}{4  (n-1) (2 n+1) \sigma ^2},
    \end{eqnarray}  
    
    therefore, it is a saddle point for
   \begin{align}
    &\alpha \leq 0\land \left(\left(2\leq n<\frac{\sigma }{\lambda }+1\land ((\sigma >0\land 0<\lambda <\sigma )\lor (\sigma <\lambda <0\land \sigma <0))\right) \right.  \nonumber \\
    &\left.\lor (n\geq 2\land ((\lambda \geq 0\land \sigma <0)\lor (\sigma >0\land \lambda \leq 0)))\right) .\nonumber \\
    &
    \end{align}

    \item Point $h$ exits for

    \begin{equation}
        (\alpha \neq 0\lor \lambda \neq 0)\land \left(\lambda =0\lor \alpha <0\lor 6 \alpha \geq \lambda ^2\right)
    \end{equation}
    
    and it has the eigenvalues
    \begin{eqnarray}
        \mu_1 &=& \frac{\lambda ^2}{\alpha }-3,\ \ \ \ \ \ \
        \mu_2 = \frac{\lambda ^2}{2 \alpha }-2, \nonumber \\
        \mu_3 &=& \frac{\lambda ^2}{2 \alpha }-3, \ \ \ \ \ \ \ 
        \mu_4 = \frac{\lambda  (\lambda +\lambda  (-n)+\sigma )}{2 \alpha },
    \end{eqnarray}  
    
    therefore, it is stable for
   \begin{align}
    &n\geq 2\land \left((\lambda =0\land \alpha \neq 0)\lor (\alpha <0\land ((\lambda <0\land \lambda +\sigma \leq \lambda  n)\lor (\lambda >0\land \lambda +\sigma \geq \lambda  n))) \right. \nonumber\\
&
\left.\lor \left(3 \alpha \geq \lambda ^2\land ((\lambda +\sigma \geq \lambda  n\land \lambda <0)\lor (\lambda >0\land \lambda +\sigma \leq \lambda  n))\right)\right). 
    \end{align}

\item Point $i$ exits for $n>1/2$ and it has the eigenvalues
    \begin{eqnarray}
        \mu_1 &=& \frac{3}{2 n-1},\ \ \ \ \ \ \
        \mu_2 = \frac{2-n}{2 n-1}, \nonumber \\
        \mu_3 &=& \frac{3 (n-1)}{2 n-1},\ \ \ \ \ \ \
        \mu_4 = \frac{3 n}{2 n-1},
    \end{eqnarray}  
    
    but, it is always unstable.
\item Point $j^{\pm}$ exits and it is stable for
    \begin{eqnarray}
        n\geq 2\land \alpha <0\land \left(\left(\lambda \geq \frac{\sigma }{n-1}\land \sigma <0\right)\lor \left(\sigma >0\land \lambda \leq \frac{\sigma }{n-1}\right)\right), \label{jj}
    \end{eqnarray}  
  however, the expressions for the eigenvalues are large and are not shown here, as they have no benefit here.

\end{itemize}

\subsection{Interaction  $Q =\beta  \kappa \rho_m \dot\phi$. Critical points and Stability analysis.}\label{inter_1}

In this subsection, we will analyze the dynamical system for our first interaction model in which the interaction term $Q$ is  defined as $Q=\kappa\beta\rho_m\dot{\phi}$ with $\beta$ an arbitrary  constant parameter. As before, we find that the different functions $f_i(x,y,u,\varrho)$ associated with the dynamical system 
are given by
\begin{eqnarray}
    f_1 &=& \frac{x}{4 n^2 u^2-2 n u^2+2 \alpha  x^2} \bigg( 
 \, 2 n^2 u^2 \left(3 u^2 + 3 \alpha x^2 - 3 y^2 + \varrho^2 + 3 \right)  - n u^2 \left(3 u^2 + 3 \alpha x^2 + 2 \sqrt{6} \sigma x - 3 y^2 + \varrho^2 + 9 \right) \nonumber \\
&& + x \left( \alpha x \left(3 u^2 - 3 y^2 + \varrho^2 - 3 \right) + \sqrt{6} \left( \sigma u^2 + \lambda y^2 \right) + 3 \alpha^2 x^3 \right) + \sqrt{6} \beta x \left((2 n-1) u^2+\alpha  x^2+y^2+\varrho ^2-1\right)
\bigg),\label{f1} \\
f_2 &=& \frac{1}{2} y \left(3 u^2+3 \alpha  x^2-\sqrt{6} \lambda  x-3 y^2+\varrho ^2+3\right), \\
f_3 &=& \frac{u}{4 n^2 u^2-2 n u^2+2 \alpha  x^2} \bigg(
 \, 2 n^2 u^2 \left(3 u^2 - 3 y^2 + 3 \alpha x^2 + \varrho^2 \right)
- n \left(3 u^4 + x \left(6 \alpha x - \sqrt{6} \lambda y^2 \right) 
+ u^2 \left(3 - 3 y^2 
+ 3 \alpha x^2 + \varrho^2 \right) \right) \nonumber \\
&& + \alpha x^2 \left(3 + 3 u^2 - 3 y^2 + 3 \alpha x^2 + \varrho^2 
+ \sqrt{6} \sigma x \right) + \sqrt{6} n \beta x \left((2 n-1) u^2+\alpha  x^2+y^2+\varrho ^2-1\right)
\bigg), \\
f_4 &=& \frac{1}{2} \varrho  \left(3 u^2+3 \alpha  x^2-3 y^2+\varrho ^2-1\right).\label{f4}
\end{eqnarray}

In this way, we study the critical points for our first interaction model, assuming the effective potential $V(\phi)$ and the coupling function $\alpha_n(\phi)=\alpha_n$ given by Eq.(\ref{VF}). As mentioned before, these critical points are determined from the conditions $dx/dN=dy/dN=du/dN=d\varrho/dN=0$ in the dynamical system given by Eqs.(\ref{dinsyseq1})-(\ref{dinsyseq8}) together with the functions defined by Eqs. (\ref{f1})-(\ref{f4}).
These critical points satisfy the following conditions $y_c\ge 0$, $u_c\ge 0$ and $\varrho_c\ge 0$, respectively. For our first interaction, these critical points are shown in Table \ref{table3} and the cosmological parameters are listed in Table \ref{table4}. 

As previously, the critical point $a_R$ represents a radiation-dominated era, in which the parameter related to the radiation $\Omega_r=1$ and the total EoS parameter $w_{tot}=1/3$. Here we note that this point does not depend on the parameters $\lambda$, $n$, and $\sigma$.

In relation to the radiation-dominated era, we note that the critical point $b_R$ corresponds to another point during this epoch, which depends on the parameters $n$, $\alpha$, and $\sigma$, in which the EoS parameters $w_{de}=w_{tot}=1/3$. We also note that this point does not depend on the interaction and  coincides with that found in  the case $Q=0$, see Table \ref{table1}. As before,  
we observe that there is a contribution associated with dark energy $\Omega_{de}$ and this contribution depends on parameters $n$, $\alpha$ and $\sigma$. Also, we note that the critical point $b_R$ only exits if the parameter $\alpha\ge0$ and $\alpha \leq \frac{(n-2) \sigma ^2}{8 (n-1)^3}$, wherewith the range for the parameter $\alpha$ is given by $\frac{(n-2) \sigma ^2}{8 (n-1)^3}\ge \alpha\ge 0$. In addition, we have a third point during the radiation-dominated epoch $c_R$, which depends on the parameters $\alpha$ and $\lambda$. As the point $b_R$, the point $c_R$ does not depend on the interaction  $Q\propto\rho_m\dot{\phi}$. However,  the point $c_R$ only exists if the parameter $\alpha$ is a positive quantity or equal to zero.

As in the case without interaction, the points $d_{SM}^{\pm}$ represent a stiff matter where the total EoS parameter $w_{tot}=1$ and $w_{de}=1$. This point does not depend on the interaction and exists only for $\alpha>0$. As before, the critical point $x_c$ has two values;  $x_c=\pm 1/\sqrt{\alpha}$ and its energy density associated with dark energy scales as $\rho_{de}\propto a^{-6}$.

In relation to the critical points $f_{M}$, $e_M$, and $g_M$, these represent matter-dominated solutions. In particular, for the critical point $e_M$ we find that the EoS parameter corresponds to dust and the matter density parameter  $\Omega_m=1$.
Also,  we note that only the point $e_M$  coincides with the situation in which there is no interaction. However, we note that the critical points $f_M$ and $g_M$  are affected by the interaction term. Here, these points depend on the parameters $\alpha$, $\lambda$, and $ \beta$, the parameter associated with the interaction.  In addition, we note that the point $f_M$ exists only if the coupling parameter $\alpha$ is positive. For this point, the density parameters $\Omega_m$ and $\Omega_{de}$ depend only  on the parameters $\alpha$ and $\lambda$. However, the EoS parameters depend on the parameters $\alpha$, $\lambda$, and $\beta$ (interaction parameter).
In contrast to point $f_M$, the point  $g_M$ exists only if the parameter $\alpha\le0$. We observe that the EoS parameters $w_{tot}$ and $w_{de}$ depend on the parameters $\alpha$, the power $n$, $\sigma$ and the interaction parameter $\beta$.

As in the case without interaction,  the critical points $h$, $i$, and $j^\pm$ are points related to the dark energy epoch, and these points do not depend on the interaction term $Q$. For each of these points,  we find that  $w_{de}=w_{tot}$ and the density parameter related to the dark energy $\Omega_{de}=1$. In particular, for the point $h$, we note that for positive values of $\alpha$, it corresponds to a quintessence-like behavior, while for negative values of  $\alpha$, the system exhibits  a phantom regime. As before, we note that the points $j^\pm$ present two values associated with the critical points $x_c$ and $u_c$, see Table \ref{table3}.

The points $k$ and $l$ correspond to two new critical points that appear due to the interaction term.  The point $k$ presents a total EoS parameter $w_{tot}=1/3$ corresponding to a radiation-dominated epoch, while the dark-energy EoS parameter is $w_{de}=1$, characteristic of stiff-matter behavior.  
The point $l$ has a total EoS parameter $w_{tot}$ that depends on the parameters  $\alpha$ and $\beta$, while the dark energy EoS parameter is $w_{de}=1$, characteristic of stiff-matter behavior, as the point $k$. Also, we note that this point presents a density parameter associated with  the radiation equal to zero.  
In addition, both points only exist if the parameter $\alpha>0$.

\begin{table*}[!b]
 \centering
 \caption{Critical points for the autonomous system in the interaction $Q\propto\rho_m \dot\phi $. With $A_c=\sigma ^2 \left(-24 \alpha  n^4+72 \alpha  n^3-72 \alpha  n^2+4 n^2 \sigma ^2+24 \alpha  n-4 n \sigma ^2+\sigma ^2\right)$}.
\begin{center}
\begin{tabular}{c c c c c c c c c}\hline\hline
Name &  $x_c$ & $y_c$ & $u_{c}$ & $\varrho_{c}$  \\\hline
$\ \ \ \ \ \ \ \ a_{R} \ \ \ \ \ \ \ \ $ & $0$ & $0$ & $0$  & $1$ \\
$\ \ \ \ \ \ \ \ b_{R} \ \ \ \ \ \ \ \ $ & $\frac{2 \sqrt{\frac{2}{3}} (n-1)}{\sigma }$ & $0$ & $\frac{2  \sqrt{\frac{2}{3}} \sqrt{\alpha } (n-1)}{\sqrt{\left((n-2) \sigma ^2\right)}}$  & $\frac{\sqrt{(n-2) \sigma ^2-8 \alpha  (n-1)^3}}{\sqrt{(n-2) \sigma ^2}}$ \\
$\ \ \ \ \ \ \ \ c_{R} \ \ \ \ \ \ \ \ $ & $\frac{2 \sqrt{\frac{2}{3}}}{\lambda }$ & $\frac{2 \sqrt{\alpha }}{\sqrt{3} \lambda }$ & $0$  & $\frac{\sqrt{\lambda ^2-4 \alpha }}{\lambda }$ \\
$\ \ \ \ \ \ \ \ d_{SM}^{\pm} \ \ \ \ \ \ \ \ $ & $\frac{\pm 1}{\sqrt{\alpha }}$ & $0$ & $0$  & $0$ \\
$\ \ \ \ \ \ \ \ e_{M} \ \ \ \ \ \ \ \ $ & $0$ & $0$ & $0$  & $0$ \\
$\ \ \ \ \ \ \ \ f_{M} \ \ \ \ \ \ \ \ $ & $\frac{\sqrt{\frac{3}{2}}}{\beta +\lambda }$ & $\frac{\sqrt{3 \alpha +2 \beta  (\beta +\lambda )}}{\sqrt{2} \sqrt{(\beta +\lambda )^2}}$ & $0$  & $0$ \\
$\ \ \ \ \ \ \ \ g_{M} \ \ \ \ \ \ \ \ $ & $\frac{\sqrt{\frac{3}{2}} (n-1)}{\beta  (n-1)+\sigma }$ & $0$ & $\frac{\sqrt{n-1} \sqrt{3 \alpha  (n-1)+2 \beta  (\beta  (n-1)+\sigma )}}{\sqrt{2} \sqrt{-(\beta  (n-1)+\sigma )^2}}$  & $0$ \\
$\ \ \ \ \ \ \ \ h \ \ \ \ \ \ \ \ $ & $\frac{\lambda }{\sqrt{6} \alpha }$ & $\sqrt{1-\frac{\lambda ^2}{6 \alpha }}$ & $0$  & $0$ \\
$\ \ \ \ \ \ \ \ i \ \ \ \ \ \ \ \ $ & $0$ & $0$ & $\frac{1}{\sqrt{2 n-1}}$  & $0$ \\
$\ \ \ \ \ \ \ \ j^{\pm} \ \ \ \ \ \ \ \ $ & $\frac{(2 n-1) \sigma ^2\pm \sqrt{A_c}}{2 \sqrt{6} \alpha  (n-1)^2 \sigma }$ & $0$ & $\frac{\sqrt{\frac{\mp\sqrt{A_c}+12 \alpha  (n-1)^3+(1-2 n) \sigma ^2}{\alpha  (n-1)^4}}}{2 \sqrt{3}}$  & $0$ \\
$\ \ \ \ \ \ \ \ k \ \ \ \ \ \ \ \ $ & $-\frac{1}{\sqrt{6} \beta }$ & $0$ & $0$  & $\frac{\sqrt{\beta ^2-\frac{\alpha }{2}}}{\beta }$ \\
$\ \ \ \ \ \ \ \ l \ \ \ \ \ \ \ \ $ & $-\frac{\sqrt{\frac{2}{3}} \beta }{\alpha }$ & $0$ & $0$  & $0$ \\

\\ \hline\hline
\end{tabular}
\end{center}
\label{table3}
\end{table*}
\begin{table}[ht]
 \centering
 \caption{Cosmological parameters for the critical points in Table \ref{table3}. With $A_c=\sigma ^2 \left(-24 \alpha  n^4+72 \alpha  n^3-72 \alpha  n^2+4 n^2 \sigma ^2+24 \alpha  n-4 n \sigma ^2+\sigma ^2\right)$.}
\begin{center}
\begin{tabular}{c c c c c c}\hline\hline
Name &   $\Omega_{de}$ & $\Omega_{m}$ & $\Omega_{r}$ & $\omega_{de}$ & $\omega_{tot}$ \\\hline
$a_{R}$ & $0$ & $0$ & $1$ & $\frac{1}{3}$ & $\frac{1}{3}$ \\
$b_{R}$ & $\frac{8 \alpha  (n-1)^3}{(n-2) \sigma ^2}$ & $0$ & $1-\frac{8 \alpha  (n-1)^3}{(n-2) \sigma ^2}$ & $\frac{1}{3}$ & $\frac{1}{3}$ \\
$c_{R}$ & $\frac{4 \alpha }{\lambda ^2}$ & $0$ & $1-\frac{4 \alpha }{\lambda ^2}$ & $\frac{1}{3}$ & $\frac{1}{3}$ \\
$d_{SM}^\pm$ & $1$ & $0$ & $0$ & $1$ & $1$ \\
$e_{M}$ & $0$ & $1$ & $0$ & $0$ & $0$ \\
$f_{M}$ & $\frac{3 \alpha }{\lambda ^2}$ & $1-\frac{3 \alpha }{\lambda ^2}$ & $0$ & $-\frac{\beta  (\beta +\lambda )}{3 \alpha +\beta  (\beta +\lambda )}$ & $-\frac{\beta }{\beta +\lambda }$ \\
$g_{M}$ & $-\frac{3 \alpha  (n-1)^3}{\sigma ^2}$ & $1+\frac{3 \alpha  (n-1)^3}{\sigma ^2}$ & $0$ & $\frac{\beta  (\beta  (n-1)+\sigma )}{3 \alpha  (n-1)^2+\beta  (2 n-1) (\beta  (n-1)+\sigma )}$ & $\frac{\beta -\beta  n}{\beta  (n-1)+\sigma }$ \\
$h$ & $1$ & $0$ & $0$ & $-1+\frac{\lambda ^2}{3 \alpha }$ & $-1+\frac{\lambda ^2}{3 \alpha }$ \\
$i$ & $1$ & $0$ & $0$ & $\frac{1}{2 n-1}$ & $\frac{1}{2 n-1}$ \\
$j^{\pm}$ & $1$ & $0$ & $0$ & $-\frac{\mp \sqrt{A_c}+6 \alpha  (n-1)^3+(1-2 n) \sigma ^2}{6 \alpha  (n-1)^3}$ & $-\frac{\mp\sqrt{A_c}+6 \alpha  (n-1)^3+(1-2 n) \sigma ^2}{6 \alpha  (n-1)^3}$ \\
$k$ & $\frac{\alpha }{6 \beta ^2}$ & $\frac{\alpha }{3 \beta ^2}$ & $1-\frac{\alpha }{2 \beta ^2}$ & $1$ & $\frac{1}{3}$ \\
$l$ & $\frac{2 \beta ^2}{3 \alpha }$ & $1-\frac{2 \beta ^2}{3 \alpha }$ & $0$ & $1$ & $\frac{2 \beta ^2}{3 \alpha }$ \\
\\ \hline\hline
\end{tabular}
\end{center}
\label{table4}
\end{table}

In the following, we provide the eigenvalues and corresponding stability conditions of the critical points for interaction  $Q =\beta  \kappa \rho_m \dot\phi$.

\begin{itemize}
    \item Point $a_R$ has the eigenvalues   
    \begin{eqnarray}
        \mu_1 &=& -1, \ \ \ \ \ \mu_2 = 1, \ \ \ \ \ 
        \ \mu_3 = 2, \ \ \ \ \ \mu_4 = 2-3n ,\nonumber\\
        &&
    \end{eqnarray}
    therefore, it is a saddle point.
    \item Point $b_R$ exits for 
    \begin{equation}
        n>2\land \alpha \leq \frac{(n-2) \sigma ^2}{8 (n-1)^3},
    \end{equation}
    and it has the eigenvalues   
   $$
        \mu_1 = 1+ \frac{4 \beta  (n-1)}{\sigma }, \ \ \ \ \ \mu_2 = \frac{2 (\lambda +\lambda  (-n)+\sigma )}{\sigma }, \ \ \ \ \ \nonumber\\
        \ \mu_3 = -\frac{\sqrt{(n+1) \sigma ^2 \left(3 (3 n-5) \sigma ^2-64 \alpha  (n-1)^3\right)}}{2 (n+1) \sigma ^2}-\frac{1}{2}, \ \ \ \ \ 
        $$
        and
       \begin{eqnarray}  
        \mu_4 = \frac{1}{2} \left(\frac{\sqrt{(n+1) \sigma ^2 \left(3 (3 n-5) \sigma ^2-64 \alpha  (n-1)^3\right)}}{(n+1) \sigma ^2}-1\right) ,\nonumber\\
        &&
    \end{eqnarray}

    therefore, it is unstable for
    \begin{eqnarray}
        &&\left(\alpha >0\land \beta >0\land 8 \alpha  (n-1)^2\leq \frac{(n-2) \sigma ^2}{n-1}\land \left(\alpha <\frac{(n-2) \sigma ^2}{8 (n-1)^3}\lor \lambda >\frac{\sigma }{n-1}\lor \sigma >0\right)\right)\\
        &&\lor \left(0<\alpha \leq \frac{(n-2) \sigma ^2}{8 (n-1)^3}\land 0<\beta <-\frac{\sigma }{4}\land \frac{\sigma }{\beta }+4 n<4\right)
    \end{eqnarray}
    \item Point $c_R$ exits for
    \begin{eqnarray}
       \left(\lambda <0\land 0\leq \alpha \leq \frac{\lambda ^2}{4}\right)\lor (\lambda =0\land \alpha =0)\lor \left(\lambda >0\land 0\leq \alpha \leq \frac{\lambda ^2}{4}\right)
    \end{eqnarray}
    it has the eigenvalues
    $$
        \mu_1 = 1+\frac{4 \beta }{\lambda }, \ \ \ \ \ \mu_2 = \frac{1}{2} \left(-\frac{\sqrt{\alpha  \lambda ^2 \left(64 \alpha -15 \lambda ^2\right)}}{\sqrt{\alpha } \lambda ^2}-1\right), \ \ \ \ \ \mu_3 = \frac{1}{2} \left(\frac{\sqrt{\alpha  \lambda ^2 \left(64 \alpha -15 \lambda ^2\right)}}{\sqrt{\alpha } \lambda ^2}-1\right), 
        $$
        and
        \begin{equation}
        \mu_4 = \frac{2 \sigma }{\lambda }-2 n+2,
    \end{equation}
    therefore, it is unstable for
    \begin{eqnarray}
        &&\left(\left(4 \alpha =\lambda ^2\lor \left(\alpha >0\land 4 \alpha \leq \lambda ^2\right)\right)\land \left(4 \alpha =\lambda ^2\lor \lambda >0\lor \lambda +\sigma \leq \lambda  n\right)\land \lambda \neq 0\land \beta >0\right)\\
        &&\lor \left(\alpha >0\land 4 \alpha \leq \lambda ^2\land \beta >0\land 4 \beta +\lambda \leq 0\right).
    \end{eqnarray}
    \item Point $d_{SM}^\pm$ exits for $\alpha >	0$ and it has the eigenvalues
    \begin{eqnarray}
        \mu_1 &=& 1, \ \ \ \ \ \mu_2 = 3, \ \ \ \ \ 
        \mu_3 = \frac{\sqrt{\frac{3}{2}} \lambda }{\sqrt{\alpha }}+3 \ \ \ \ \ 
        \mu_4 = -\frac{\sqrt{\frac{3}{2}} \sigma }{\sqrt{\alpha }}-3 n+3,
    \end{eqnarray}
    therefore, it is unstable.
    \item Point $e_M$ has the eigenvalues   
    \begin{eqnarray}
        \mu_1 &=& -\frac{1}{2}, \ \ \ \ \ \mu_2 = -\frac{3}{2}, \ \ \ \ \ 
        \ \mu_3 = \frac{3}{2}, \ \ \ \ \ \mu_4 = \frac{1}{2} (3-6 n) ,\nonumber\\
        &&
    \end{eqnarray}
    therefore, it is a saddle point.
    \item Point $f_{M}$ exits for $\alpha \geq 0$ and $n > 2$, and it has the eigenvalues
    $$
        \mu_1 = -\frac{4 \beta +\lambda }{2 (\beta +\lambda )},\ \ \ \ \ \ \
        \mu_2 = \frac{\sqrt{3} \sqrt{-\alpha  (\beta +\lambda )^8 \left(72 \alpha ^2+3 \alpha  \left(20 \beta ^2+12 \beta  \lambda -7 \lambda ^2\right)-16 \beta  \lambda  (\beta +\lambda )^2\right)}}{4 \alpha  (\beta +\lambda )^4 \sqrt{-(\beta +\lambda )^2}}-\frac{3 (2 \beta +\lambda )}{4 (\beta +\lambda )}, \nonumber \\
    $$
    $$
    \mu_3 = \frac{3 (\lambda +\lambda  (-n)+\sigma )}{2 (\beta +\lambda )}, 
        $$
        and
       \begin{eqnarray}   
        \mu_4 = -\frac{\sqrt{3} \sqrt{-\alpha  (\beta +\lambda )^8 \left(72 \alpha ^2+3 \alpha  \left(20 \beta ^2+12 \beta  \lambda -7 \lambda ^2\right)-16 \beta  \lambda  (\beta +\lambda )^2\right)}}{4 \alpha  \left(-(\beta +\lambda )^2\right)^{5/2}}-\frac{3 (2 \beta +\lambda )}{4 (\beta +\lambda )},\nonumber\\
        &&
    \end{eqnarray}

    therefore, it is a saddle point for
$$   
(\lambda <0\land ((3 \alpha =\lambda  (\beta +\lambda )\land 2 \beta +\lambda =0)\lor (3 \alpha +2 \beta  (\beta +\lambda )\geq 0\land 2 \beta +\lambda <0\land 3 \alpha \leq \lambda  (\beta +\lambda )\land 
$$
\begin{eqnarray}
((\beta >0\land \lambda +\sigma \leq \lambda  n)\lor 4 \beta +\lambda \geq 0))))\lor (\alpha >0\land \lambda +\sigma \geq \lambda  n\land 3 \alpha \leq \lambda  (\beta +\lambda )\land \beta >0\land \lambda >0).
   \end{eqnarray}

    \item Point $g_{M}$ exits for 
    \begin{equation}
        \beta >0\land n\geq 2\land \alpha <\frac{2 \beta ^2-2 \beta  \sigma -2 \beta ^2 n}{3 n-3},
    \end{equation}
    and it has the eigenvalues
    \begin{eqnarray}
        \mu_1 &=& -\frac{1}{2},\ \ \ \ \ \ \
        \mu_2 = \frac{3 (\lambda +\lambda  (-n)+\sigma )}{2 \sigma }, \nonumber \\
        \mu_3 &=& -\frac{3 \left( \left(2 n^2-n-1\right) \sigma ^2+\sqrt{ \left(-(n-1)^2\right) (2 n+1) \sigma ^2 \left(24 \alpha  (n-1)^3+(7-2 n) \sigma ^2\right)}\right)}{4  (n-1) (2 n+1) \sigma ^2},\nonumber \\
        \mu_4 &=& \frac{3 \left( \left(-2 n^2+n+1\right) \sigma ^2+\sqrt{  \left(-(n-1)^2\right) (2 n+1) \sigma ^2 \left(24 \alpha  (n-1)^3+(7-2 n) \sigma ^2\right)}\right)}{4  (n-1) (2 n+1) \sigma ^2},
    \end{eqnarray}  
    
    therefore, it is a saddle point for
   \begin{align}
    &\alpha \leq 0\land \left(\left(2\leq n<\frac{\sigma }{\lambda }+1\land ((\sigma >0\land 0<\lambda <\sigma )\lor (\sigma <\lambda <0\land \sigma <0))\right) \right.  \nonumber \\
    &\left.\lor (n\geq 2\land ((\lambda \geq 0\land \sigma <0)\lor (\sigma >0\land \lambda \leq 0)))\right) .\nonumber \\
    &
    \end{align}

    \item Point $h$ exits for

    \begin{equation}
        (\alpha \neq 0\lor \lambda \neq 0)\land \left(\lambda =0\lor \alpha <0\lor 6 \alpha \geq \lambda ^2\right)
    \end{equation}
    
    and it has the eigenvalues
    \begin{eqnarray}
        \mu_1 &=& \frac{\lambda ^2}{\alpha }-3,\ \ \ \ \ \ \
        \mu_2 = \frac{\lambda ^2}{2 \alpha }-2, \nonumber \\
        \mu_3 &=& \frac{\lambda ^2}{2 \alpha }-3, \ \ \ \ \ \ \ 
        \mu_4 = \frac{\lambda  (\lambda +\lambda  (-n)+\sigma )}{2 \alpha },
    \end{eqnarray}  
    
    therefore, it is stable for
   \begin{align}
    &n\geq 2\land \left((\lambda =0\land \alpha \neq 0)\lor (\alpha <0\land ((\lambda <0\land \lambda +\sigma \leq \lambda  n)\lor (\lambda >0\land \lambda +\sigma \geq \lambda  n))) \right. \nonumber\\
&
\left.\lor \left(3 \alpha \geq \lambda ^2\land ((\lambda +\sigma \geq \lambda  n\land \lambda <0)\lor (\lambda >0\land \lambda +\sigma \leq \lambda  n))\right)\right). 
    \end{align}

\item Point $i$ exits for $n>1/2$ and it has the eigenvalues
    \begin{eqnarray}
        \mu_1 &=& \frac{3}{2 n-1},\ \ \ \ \ \ \
        \mu_2 = \frac{2-n}{2 n-1}, \nonumber \\
        \mu_3 &=& \frac{3 (n-1)}{2 n-1},\ \ \ \ \ \ \
        \mu_4 = \frac{3 n}{2 n-1},
    \end{eqnarray}  
    
    but, it is always unstable.
\item Point $j^{\pm}$ exits and it is stable for
    \begin{eqnarray}
        n\geq 2\land \alpha <0\land \left(\left(\lambda \geq \frac{\sigma }{n-1}\land \sigma <0\right)\lor \left(\sigma >0\land \lambda \leq \frac{\sigma }{n-1}\right)\right), 
    \end{eqnarray} 
  however, the expressions for the eigenvalues are large and are not shown here, as they have no benefit here.
  \item Points $k$ and $l$ do not play a significant role in the cosmological evolution and will therefore not be considered in the stability analysis, since the EoS parameters $w_{de}$ and $w_{total}$ are positive quantities and represent similar behavior to the stiff matter.

\end{itemize}

\subsection{Interaction  $Q \propto \rho_m H$. Critical points and Stability analysis.}\label{inter_2}

In this subsection, we will analyze the dynamical system for our second interaction model defined as $Q=3 \beta \rho_m H$ with $\beta$ a constant parameter. As before, we find that the different functions $f_i(x,y,u,\varrho)$ associated with the dynamical system 
are given by
\begin{eqnarray}
    f_1 &=& \frac{x}{4 n^2 u^2-2 n u^2+2 \alpha  x^2} \bigg( 
 \, 2 n^2 u^2 \left(3 u^2 + 3 \alpha x^2 - 3 y^2 + \varrho^2 + 3 \right)  - n u^2 \left(3 u^2 + 3 \alpha x^2 + 2 \sqrt{6} \sigma x - 3 y^2 + \varrho^2 + 9 \right) \nonumber \\
&& + x \left( \alpha x \left(3 u^2 - 3 y^2 + \varrho^2 - 3 \right) + \sqrt{6} \left( \sigma u^2 + \lambda y^2 \right) + 3 \alpha^2 x^3 \right) + 3 \beta \left((2 n-1) u^2+\alpha  x^2+y^2+\varrho ^2-1\right)
\bigg),\label{f12} \\
f_2 &=& \frac{1}{2} y \left(3 u^2+3 \alpha  x^2-\sqrt{6} \lambda  x-3 y^2+\varrho ^2+3\right), \\
f_3 &=& \frac{u}{4 n^2 u^2-2 n u^2+2 \alpha  x^2} \bigg(
 \, 2 n^2 u^2 \left(3 u^2 - 3 y^2 + 3 \alpha x^2 + \varrho^2 \right)
- n \left(3 u^4 + x \left(6 \alpha x - \sqrt{6} \lambda y^2 \right) 
+ u^2 \left(3 - 3 y^2 + 3 \alpha x^2 + \varrho^2 \right) \right) \nonumber \\
&& + \alpha x^2 \left(3 + 3 u^2 - 3 y^2 + 3 \alpha x^2 + \varrho^2 
+ \sqrt{6} \sigma x \right) + 3 n \beta \left((2 n-1) u^2+\alpha  x^2+y^2+\varrho ^2-1\right)
\bigg), \\
f_4 &=& \frac{1}{2} \varrho  \left(3 u^2+3 \alpha  x^2-3 y^2+\varrho ^2-1\right).\label{f42}
\end{eqnarray}

As before, we study the critical points for our second interaction model, assuming the effective potential $V(\phi)$ and the coupling function $\alpha_n(\phi)=\alpha_n$ given by Eq.(\ref{VF}). 
In the following, we provide the eigenvalues and the corresponding stability conditions of the critical points for interaction  $Q = 3 \beta \rho_m H$.

\begin{table*}[ht]
 \centering
 \caption{Critical points for the autonomous system in the interaction $Q\propto\rho_m H $. With $A_c=\sigma ^2 \left(-24 \alpha  n^4+72 \alpha  n^3-72 \alpha  n^2+4 n^2 \sigma ^2+24 \alpha  n-4 n \sigma ^2+\sigma ^2\right)$}.
\begin{center}
\begin{tabular}{c c c c c c c c c}\hline\hline
Name &  $x_c$ & $y_c$ & $u_{c}$ & $\varrho_{c}$  \\\hline
$\ \ \ \ \ \ \ \ a_{R} \ \ \ \ \ \ \ \ $ & $0$ & $0$ & $0$  & $1$ \\
$\ \ \ \ \ \ \ \ b_{R} \ \ \ \ \ \ \ \ $ & $\frac{2 \sqrt{\frac{2}{3}} (n-1)}{\sigma }$ & $0$ & $\frac{2  \sqrt{\frac{2}{3}} \sqrt{\alpha } (n-1)}{\sqrt{\left((n-2) \sigma ^2\right)}}$  & $\frac{\sqrt{(n-2) \sigma ^2-8 \alpha  (n-1)^3}}{\sqrt{(n-2) \sigma ^2}}$ \\
$\ \ \ \ \ \ \ \ c_{R} \ \ \ \ \ \ \ \ $ & $\frac{2 \sqrt{\frac{2}{3}}}{\lambda }$ & $\frac{2 \sqrt{\alpha }}{\sqrt{3} \lambda }$ & $0$  & $\frac{\sqrt{\lambda ^2-4 \alpha }}{\lambda }$ \\
$\ \ \ \ \ \ \ \ d_{SM}^{\pm} \ \ \ \ \ \ \ \ $ & $\frac{\pm 1}{\sqrt{\alpha }}$ & $0$ & $0$  & $0$ \\
$\ \ \ \ \ \ \ \ e_{M} \ \ \ \ \ \ \ \ $ & $0$ & $0$ & $0$  & $0$ \\
$\ \ \ \ \ \ \ \ f_{M} \ \ \ \ \ \ \ \ $ & $\frac{\sqrt{\frac{3}{2}} (1-\beta)}{\lambda }$ & $\frac{\sqrt{3 \alpha  (\beta -1)^2+2 \beta  \lambda ^2}}{\sqrt{2} \lambda }$ & $0$  & $0$ \\
$\ \ \ \ \ \ \ \ g_{M} \ \ \ \ \ \ \ \ $ & $-\frac{\sqrt{\frac{3}{2}} (\beta -1) (n-1)}{\sigma }$ & $0$ & $\frac{\sqrt{-2 \beta  \sigma ^2-3 \alpha  (\beta -1)^2 (n-1)^2}}{\sqrt{2} \sigma }$  & $0$ \\
$\ \ \ \ \ \ \ \ h \ \ \ \ \ \ \ \ $ & $\frac{\lambda }{\sqrt{6} \alpha }$ & $\sqrt{1-\frac{\lambda ^2}{6 \alpha }}$ & $0$  & $0$ \\
$\ \ \ \ \ \ \ \ i \ \ \ \ \ \ \ \ $ & $0$ & $0$ & $\frac{1}{\sqrt{2 n-1}}$  & $0$ \\
$\ \ \ \ \ \ \ \ j^{\pm} \ \ \ \ \ \ \ \ $ & $\frac{(2 n-1) \sigma ^2\pm \sqrt{A_c}}{2 \sqrt{6} \alpha  (n-1)^2 \sigma }$ & $0$ & $\frac{\sqrt{\frac{\mp\sqrt{A_c}+12 \alpha  (n-1)^3+(1-2 n) \sigma ^2}{\alpha  (n-1)^4}}}{2 \sqrt{3}}$  & $0$ \\
$\ \ \ \ \ \ \ \ k \ \ \ \ \ \ \ \ $ & $0$ & $0$ & $\sqrt{-\beta}$  & $0$ \\
$\ \ \ \ \ \ \ \ l \ \ \ \ \ \ \ \ $ & $\frac{\sqrt{-\beta }}{\sqrt{\alpha }}$ & $0$ & $0$  & $0$ \\

\\ \hline\hline
\end{tabular}
\end{center}
\label{table5}
\end{table*}
\begin{table}[ht]
 \centering
 \caption{Cosmological parameters for the critical points in Table \ref{table5}. With $A_c=\sigma ^2 \left(-24 \alpha  n^4+72 \alpha  n^3-72 \alpha  n^2+4 n^2 \sigma ^2+24 \alpha  n-4 n \sigma ^2+\sigma ^2\right)$.}
\begin{center}
\begin{tabular}{c c c c c c}\hline\hline
Name &   $\Omega_{de}$ & $\Omega_{m}$ & $\Omega_{r}$ & $\omega_{de}$ & $\omega_{tot}$ \\\hline
$a_{R}$ & $0$ & $0$ & $1$ & $\frac{1}{3}$ & $\frac{1}{3}$ \\
$b_{R}$ & $\frac{8 \alpha  (n-1)^3}{(n-2) \sigma ^2}$ & $0$ & $1-\frac{8 \alpha  (n-1)^3}{(n-2) \sigma ^2}$ & $\frac{1}{3}$ & $\frac{1}{3}$ \\
$c_{R}$ & $\frac{4 \alpha }{\lambda ^2}$ & $0$ & $1-\frac{4 \alpha }{\lambda ^2}$ & $\frac{1}{3}$ & $\frac{1}{3}$ \\
$d_{SM}^\pm$ & $1$ & $0$ & $0$ & $1$ & $1$ \\
$e_{M}$ & $0$ & $1$ & $0$ & $0$ & $0$ \\
$f_{M}$ & $\frac{3 \alpha  (\beta -1)^2+\beta  \lambda ^2}{\lambda ^2}$ & $1-\frac{3 \alpha  (\beta -1)^2+\beta  \lambda ^2}{\lambda ^2}$ & $0$ & $-\frac{\beta  \lambda ^2}{3 \alpha  (\beta -1)^2+\beta  \lambda ^2}$ & $-\beta$ \\
$g_{M}$ & $-\frac{3 \alpha  (\beta -1)^2 (n-1)^3+ \beta \sigma^2  (2 n-1)}{\sigma ^2}$ & $1+\frac{3 \alpha  (\beta -1)^2 (n-1)^3+ \beta \sigma^2  (2 n-1)}{\sigma ^2}$ & $0$  & $\frac{\beta  \sigma ^2}{3 \alpha  (\beta -1)^2 (n-1)^3+\beta  (2 n-1) \sigma ^2}$ & $-\beta$\\
$h$ & $1$ & $0$ & $0$ & $-1+\frac{\lambda ^2}{3 \alpha }$ & $-1+\frac{\lambda ^2}{3 \alpha }$ \\
$i$ & $1$ & $0$ & $0$ & $\frac{1}{2 n-1}$ & $\frac{1}{2 n-1}$ \\
$j^{\pm}$ & $1$ & $0$ & $0$ & $-\frac{\mp \sqrt{A_c}+6 \alpha  (n-1)^3+(1-2 n) \sigma ^2}{6 \alpha  (n-1)^3}$ & $-\frac{\mp\sqrt{A_c}+6 \alpha  (n-1)^3+(1-2 n) \sigma ^2}{6 \alpha  (n-1)^3}$ \\
$k$ & $-\beta  (2 n-1)$ & $1+\beta  (2 n-1)$ & $0$ & $\frac{1}{2 n-1}$ & $-\beta$ \\
$l$ & $-\beta$ & $1+\beta$ & $0$ & $1$ & $-\beta$ \\
\\ \hline\hline
\end{tabular}
\end{center}
\label{table6}
\end{table}

\begin{itemize}
    \item Point $a_R$ has the eigenvalues   
    \begin{eqnarray}
        \mu_1 &=& -1, \ \ \ \ \ \mu_2 = 1, \ \ \ \ \ 
        \ \mu_3 = 2, \ \ \ \ \ \mu_4 = 2-3n ,\nonumber\\
        &&
    \end{eqnarray}
    therefore, it is a saddle point.
    \item Point $b_R$ exits for 
    \begin{equation}
        n>2\land \alpha \leq \frac{(n-2) \sigma ^2}{8 (n-1)^3},
    \end{equation}
    and it has the eigenvalues   
    $$
        \mu_1 = 1+ \frac{4 \beta  (n-1)}{\sigma }, \ \ \ \ \ \mu_2 = \frac{2 (\lambda +\lambda  (-n)+\sigma )}{\sigma }, \ \ \ \ \ \nonumber\\
        \ \mu_3 = -\frac{\sqrt{(n+1) \sigma ^2 \left(3 (3 n-5) \sigma ^2-64 \alpha  (n-1)^3\right)}}{2 (n+1) \sigma ^2}-\frac{1}{2}, 
        $$
        and
       \begin{eqnarray} 
        \mu_4 = \frac{1}{2} \left(\frac{\sqrt{(n+1) \sigma ^2 \left(3 (3 n-5) \sigma ^2-64 \alpha  (n-1)^3\right)}}{(n+1) \sigma ^2}-1\right) ,
    \end{eqnarray}

    therefore, it is unstable for
    \begin{eqnarray}
        &&\left(\alpha >0\land \beta >0\land 8 \alpha  (n-1)^2\leq \frac{(n-2) \sigma ^2}{n-1}\land \left(\alpha <\frac{(n-2) \sigma ^2}{8 (n-1)^3}\lor \lambda >\frac{\sigma }{n-1}\lor \sigma >0\right)\right)\\
        &&\lor \left(0<\alpha \leq \frac{(n-2) \sigma ^2}{8 (n-1)^3}\land 0<\beta <-\frac{\sigma }{4}\land \frac{\sigma }{\beta }+4 n<4\right).
    \end{eqnarray}
    \item Point $c_R$ exits for
    \begin{eqnarray}
       \left(\lambda <0\land 0\leq \alpha \leq \frac{\lambda ^2}{4}\right)\lor (\lambda =0\land \alpha =0)\lor \left(\lambda >0\land 0\leq \alpha \leq \frac{\lambda ^2}{4}\right)
    \end{eqnarray}
    it has the eigenvalues
    $$
        \mu_1 = 1+\frac{4 \beta }{\lambda }, \ \ \ \ \ \mu_2 = \frac{1}{2} \left(-\frac{\sqrt{\alpha  \lambda ^2 \left(64 \alpha -15 \lambda ^2\right)}}{\sqrt{\alpha } \lambda ^2}-1\right), \ \ \ \ \ \mu_3 = \frac{1}{2} \left(\frac{\sqrt{\alpha  \lambda ^2 \left(64 \alpha -15 \lambda ^2\right)}}{\sqrt{\alpha } \lambda ^2}-1\right),
        $$
        and
        \begin{equation}
        \ \ \ \ \ \mu_4 = \frac{2 \sigma }{\lambda }-2 n+2,
    \end{equation}
    therefore, it is unstable for
    \begin{eqnarray}
        &&\left(\left(4 \alpha =\lambda ^2\lor \left(\alpha >0\land 4 \alpha \leq \lambda ^2\right)\right)\land \left(4 \alpha =\lambda ^2\lor \lambda >0\lor \lambda +\sigma \leq \lambda  n\right)\land \lambda \neq 0\land \beta >0\right)\\
        &&\lor \left(\alpha >0\land 4 \alpha \leq \lambda ^2\land \beta >0\land 4 \beta +\lambda \leq 0\right).
    \end{eqnarray}
    \item Point $d_{SM}^\pm$ exits for $\alpha >	0$ and it has the eigenvalues
    \begin{eqnarray}
        \mu_1 &=& 1, \ \ \ \ \ \mu_2 = 3, \ \ \ \ \ 
        \mu_3 = \frac{\sqrt{\frac{3}{2}} \lambda }{\sqrt{\alpha }}+3 \ \ \ \ \ 
        \mu_4 = -\frac{\sqrt{\frac{3}{2}} \sigma }{\sqrt{\alpha }}-3 n+3,
    \end{eqnarray}
    therefore, it is unstable.
    \item Point $e_M$ has the eigenvalues   
    \begin{eqnarray}
        \mu_1 &=& -\frac{1}{2}, \ \ \ \ \ \mu_2 = -\frac{3}{2}, \ \ \ \ \ 
        \ \mu_3 = \frac{3}{2}, \ \ \ \ \ \mu_4 = \frac{1}{2} (3-6 n) ,\nonumber\\
        &&
    \end{eqnarray}
    therefore, it is a saddle point.
    \item Point $f_{M}$  has the eigenvalues
    \begin{eqnarray}
        \mu_1 &=& \frac{-1}{2} (3 \beta +1), \ \ \mu_2 = \frac{3 (\beta -1) (\lambda  (n-1)-\sigma )}{2 \lambda }\\
        \mu_3 &=& -\frac{\sqrt{f_\lambda}+3 \alpha  \left(3 \beta ^2-2 \beta -1\right) \lambda ^2+2 \beta  \lambda ^4}{4 \alpha  (\beta -1) \lambda ^2} \\
        , \ \ \ \ \ \ \ 
        \mu_4 &=& \frac{\sqrt{f_\lambda}-9 \alpha  \beta ^2 \lambda ^2+6 \alpha  \beta  \lambda ^2+3 \alpha  \lambda ^2-2 \beta  \lambda ^4}{4 \alpha  (\beta -1) \lambda ^2},\nonumber\\
        &&
    \end{eqnarray}  
    where, $f_\lambda= -216 \alpha ^3 (\beta -1)^5 \lambda ^2-9 \alpha ^2 (\beta -1)^2 \left(15 \beta ^2-38 \beta +7\right) \lambda ^4-12 \alpha  \beta  \left(\beta ^2-6 \beta +5\right) \lambda ^6+4 \beta ^2 \lambda ^8$. Therefore, it is a saddle point for
$$   
(\lambda <0\land ((3 \alpha =\lambda  (\beta +\lambda )\land 2 \beta +\lambda =0)\lor (3 \alpha +2 \beta  (\beta +\lambda )\geq 0\land 2 \beta +\lambda <0\land 3 \alpha \leq \lambda  (\beta +\lambda )\land 
$$
\begin{eqnarray}
((\beta >0\land \lambda +\sigma \leq \lambda  n)\lor 4 \beta +\lambda \geq 0))))
\lor (\alpha >0\land \lambda +\sigma \geq \lambda  n\land 3 \alpha \leq \lambda  (\beta +\lambda )\land \beta >0\land \lambda >0)
   \end{eqnarray}

    \item Point $g_{M}$ exits and it is saddle point for
   \begin{align}
    &\left(\left(2\leq n<\frac{\sigma }{\lambda }+1\land ((\sigma >0\land 0<\lambda <\sigma )\lor (\sigma <\lambda <0\land \sigma <0))\right) \right.  \nonumber \\
    &\left.\lor (n\geq 2\land ((\lambda \geq 0\land \sigma <0)\lor (\sigma >0\land \lambda \leq 0)))\right) ,\nonumber \\
    &
    \end{align}
however, the expressions for the eigenvalues are large and are not shown here, as they have no benefit here.
    \item Point $h$ exits for

    \begin{equation}
        (\alpha \neq 0\lor \lambda \neq 0)\land \left(\lambda =0\lor \alpha <0\lor 6 \alpha \geq \lambda ^2\right)
    \end{equation}
    
    and it has the eigenvalues
    \begin{eqnarray}
        \mu_1 &=& \frac{\lambda ^2}{\alpha }-3,\ \ \ \ \ \ \
        \mu_2 = \frac{\lambda ^2}{2 \alpha }-2, \nonumber \\
        \mu_3 &=& \frac{\lambda ^2}{2 \alpha }-3, \ \ \ \ \ \ \ 
        \mu_4 = \frac{\lambda  (\lambda +\lambda  (-n)+\sigma )}{2 \alpha },
    \end{eqnarray}  
    
    therefore, it is stable for
   \begin{align}
    &n\geq 2\land \left((\lambda =0\land \alpha \neq 0)\lor (\alpha <0\land ((\lambda <0\land \lambda +\sigma \leq \lambda  n)\lor (\lambda >0\land \lambda +\sigma \geq \lambda  n))) \right. \nonumber\\
&
\left.\lor \left(3 \alpha \geq \lambda ^2\land ((\lambda +\sigma \geq \lambda  n\land \lambda <0)\lor (\lambda >0\land \lambda +\sigma \leq \lambda  n))\right)\right). 
    \end{align}

\item Point $i$ exits for $n>1/2$ and it has the eigenvalues
    \begin{eqnarray}
        \mu_1 &=& \frac{3}{2 n-1},\ \ \ \ \ \ \
        \mu_2 = \frac{2-n}{2 n-1}, \nonumber \\
        \mu_3 &=& \frac{3 (n-1)}{2 n-1},\ \ \ \ \ \ \
        \mu_4 = \frac{3 n}{2 n-1},
    \end{eqnarray}  
    
    but, it is always unstable.
\item Point $j^{\pm}$ exits and it is stable for
    \begin{eqnarray}
        n\geq 2\land \alpha <0\land \left(\left(\lambda \geq \frac{\sigma }{n-1}\land \sigma <0\right)\lor \left(\sigma >0\land \lambda \leq \frac{\sigma }{n-1}\right)\right), 
    \end{eqnarray} 
  however, the expressions for the eigenvalues are large and are not shown here, as they have no benefit here.
  \item Points $k$ and $l$ do not play a significant role in the cosmological evolution and will therefore not be considered in the stability analysis. As before,in these points,   the EoS parameters $w_{de}$ and $w_{total}$ are positive quantities and represent similar behavior to the stiff matter.
    
\end{itemize}

In relation to the critical points, as mentioned earlier, the critical point $a _R$ is associated with a radiation-dominated era in which the density parameter $\Omega_r=1$ and both the total and dark energy equation of state parameters take the value $1/3$. As before, the critical point $b_R$ corresponds to the other point during the radiation-dominated epoch, and this point exists only for positive values of $\alpha$, and it depends on the parameter $\sigma$ and the power $n$, with the EoS parameters $w_{de}=w_{tot}=1/3$. As before, we note that the point $b_R$ does not depend on the interaction and coincides with the point obtained in the situation in which $Q=0$. Also, we note that for large values of the parameter $\sigma$, the density parameter related to the radiation $\Omega_r\rightarrow 1$, while  $\Omega_{de}\rightarrow 0$. The critical point $c_R$ corresponds to another point during the radiation-dominated epoch where the EoS parameters of $w_{de}=w_{tot}=1/3$. In addition,   this point depends on the parameters $\lambda$ and $\alpha$, with $\alpha>0$, and moreover, this point becomes independent of the power $n$ and the interaction parameter $\beta$.   In relation to the points $d_{SM}^{\pm}$, these represent a stiff matter where both the total and dark energy EoS parameters are $w_{tot}=w_{de}=1$. These points do not depend on the interaction and exist only for $\alpha>0$. As before, the critical point $x_c$ has two values;  $x_c=\pm 1/\sqrt{\alpha}$ and its energy density related with dark energy scales as $\rho_{de}\propto a^{-6}$.

Now, for the critical points $e_{M}$, $f_M$, and $g_M$, these correspond to matter-dominated solutions. We note that only for the point $e_M$ does the EoS parameter correspond to dust,  and then $\Omega_m=1$.
Additionally,  we observe that only the point $e_M$  coincides with the case in which there is no interaction, that is,  when $Q=0$. For
 the critical points $f_M$ and $g_M$, we note that these points are affected by the interaction term, since they incorporate the parameter $\beta$. In addition, 
 these points 
 depend on the parameters $\alpha$, $\lambda$.  Moreover, the point $f_M$ exists only if the coupling parameter $\alpha$ is positive. Thus, for this critical point the density parameters $\Omega_m$ and $\Omega_{de}$ depend   on the parameters $\alpha$, $\lambda$ and the interaction parameter $\beta$. Similarly, the EoS parameter $w_{de}$ depends on the parameter $\alpha$, $\lambda$ and $\beta$ and $w_{tot}$ only depends on the interaction parameter $\beta$.
In contrast to point $f_M$, the point  $g_M$ may exist if the parameter $\alpha\le0$.  For this critical point, we observe that the EoS parameter  $w_{de}$ depends on the parameters $\alpha$, the power $n$, $\sigma$, and the interaction parameter $\beta$, and the total EoS parameter $w_{tot}$ only on $\beta$. Also, we note that for large values of the parameter $\sigma$ associated with the coupling parameter $\alpha_n(\phi)$, the dark energy density parameter $\Omega_{de}\to 0$ and $\Omega_{m}\to 1$.

As before,  the critical points $h$, $i$, and $j^\pm$ are points related to the dark energy epoch in which the dark energy parameter $\Omega_{de}=1$, and these points do not depend on the interaction term related to $Q$. For each of these points,  we note that  $w_{de}=w_{tot}$ and the density parameters related to the radiation   and matter are zero. In particular, for the point $h$, we note that for positive values of $\alpha$ it corresponds to a quintessence-like behavior, while $\alpha<0$  corresponds to a phantom regime. As in the previous cases, we have that the points $j^\pm$ exhibit  two values related to the critical points $x_c$ and $u_c$, respectively.

As in the case of the first interaction, the points $k$ and $l$ are two new critical points that appear due to the interaction term.  Both points exhibit a total EoS parameter $w_{tot}=-\beta$ and correspond to a matter-dominated epoch in the limit $\beta\to 0$.  In addition, both points only exist if the parameter $\alpha>0$. However,  the points $k$ and $l$ do not play a significant role in the cosmological evolution of the system and are not associated with any physically relevant epoch; consequently, they will not be considered in the stability analysis.

\begin{figure}[h!]
\centering
\begin{minipage}{0.4\textwidth}
    \centering
    \includegraphics[width=\linewidth]{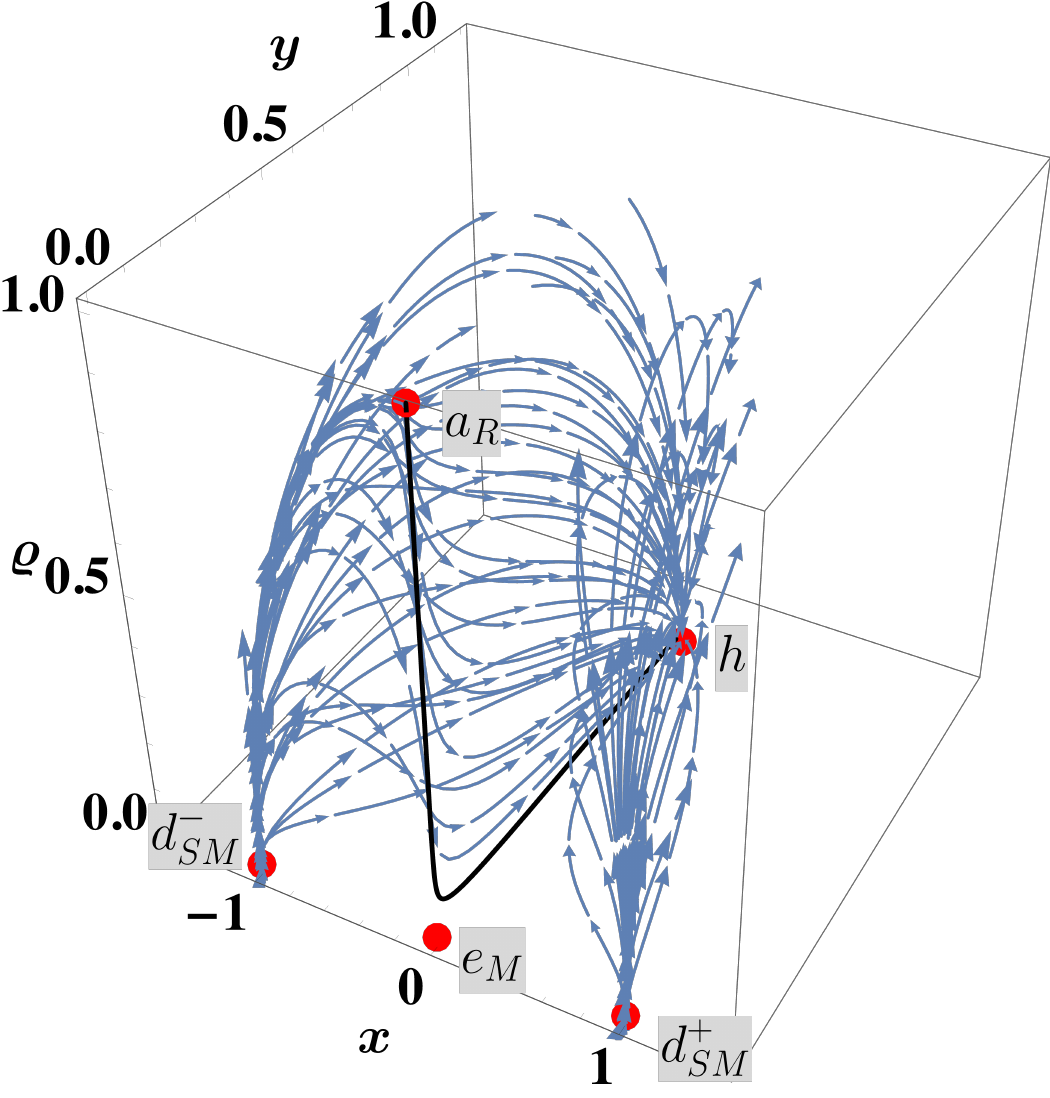}
\end{minipage}
\hfill
\begin{minipage}{0.4\textwidth}
    \centering
    \includegraphics[width=\linewidth]{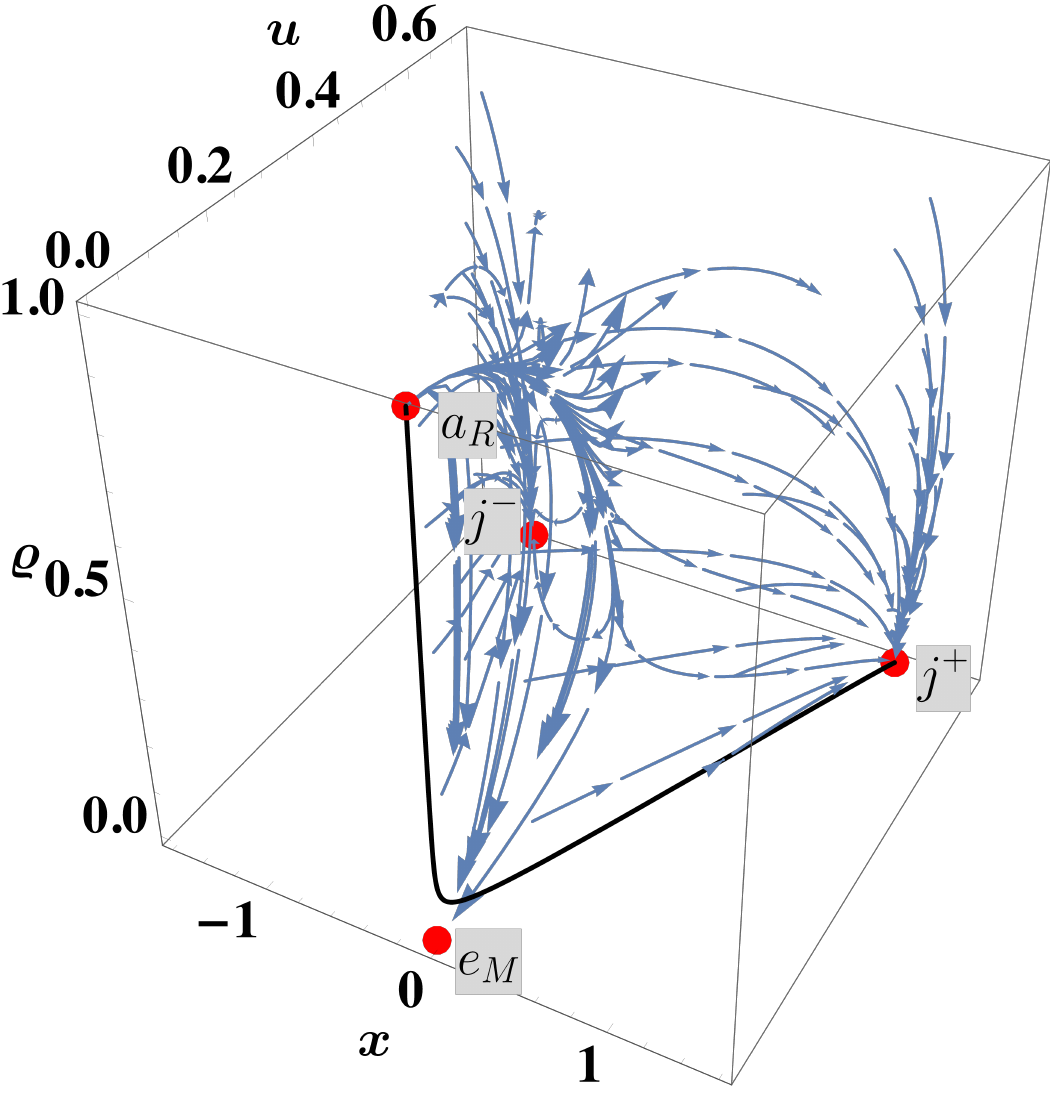}
\end{minipage}
\caption{ Phase-space evolutions for our non-interacting model, for two different values of  $\alpha$, when the parameter $n=3$. The left panel shows the trajectories in the phase space for the case in which the parameter $\alpha=+1$, specifically the paths $a_R\to e_M$ and the attractor $h$. In particular, the black curve corresponds to the initial condition $x_i= 10 \times 10^{-6}$, $y_i=8.54 \times 10^{-13}$, $u_i=5 \times 10^{-5} $ and $\varrho_i=0.99983$. The right panel illustrates the trajectories in the phase space for the case where the parameter is negative, $\alpha=-1$. Here, we show the paths $a_R\to e_M$ and the attractors are  $j^{\pm}$. In this case, the black curve corresponds to the initial condition $x_i= 10 \times 10^{-12}$, $y_i=5.1 \times 10^{-15}$, $u_i=1.179 \times 10^{-12} $ and $\varrho_i=0.99983$
}
\label{Fig1-1}
\end{figure}

\begin{figure}[h!]
\centering
\begin{tabular}{cc}
\includegraphics[width=0.45\textwidth]{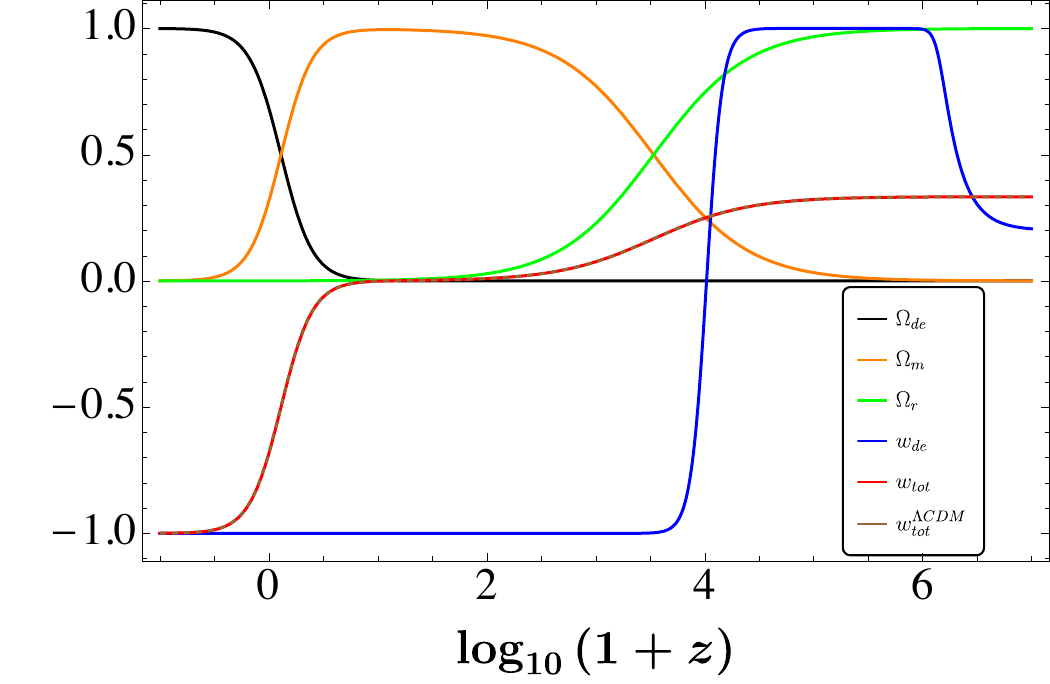} &
\includegraphics[width=0.45\textwidth]{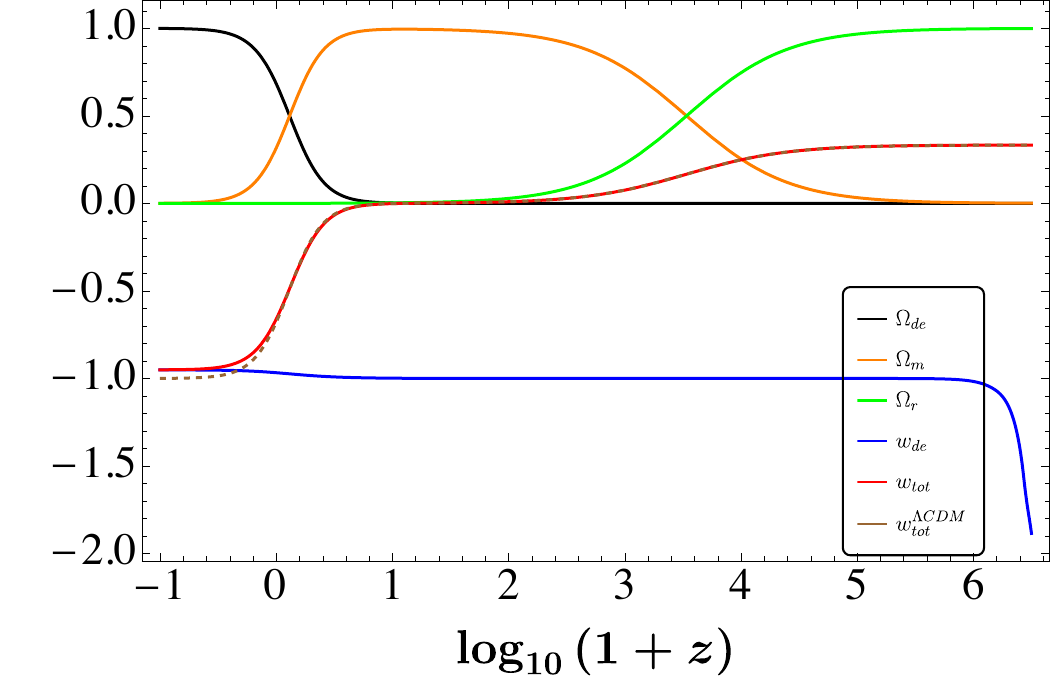} \\
\includegraphics[width=0.45\textwidth]{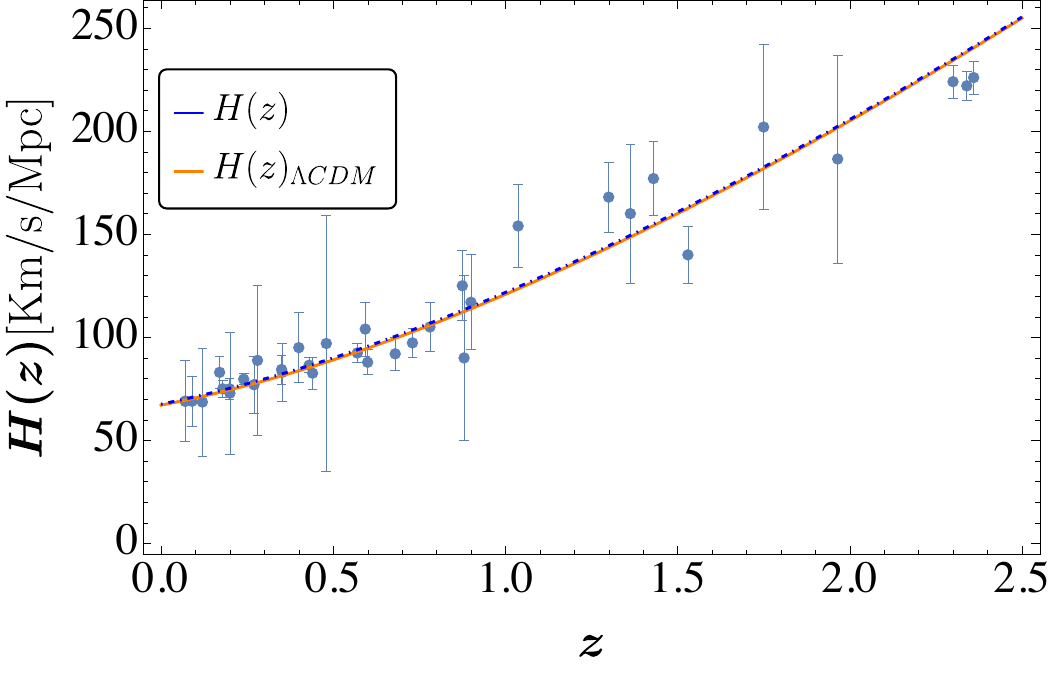} &
\includegraphics[width=0.45\textwidth]{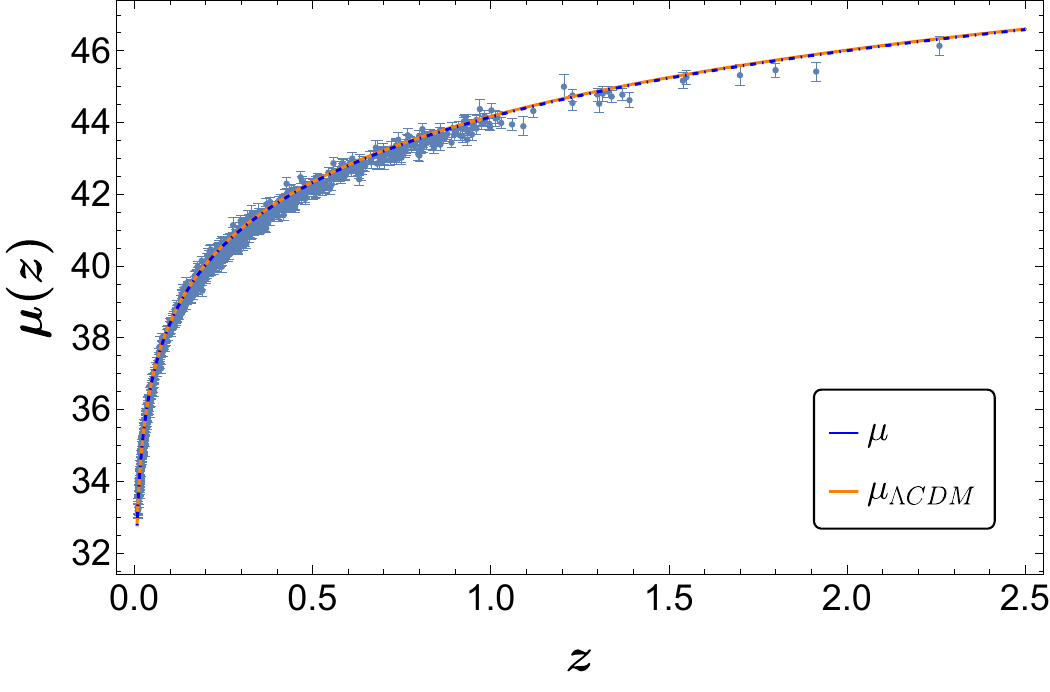}
\end{tabular}
\caption{Non-interacting model: The upper-left panel shows the evolution of the fractional energy densities of dark energy, matter, and radiation, together with the equation-of-state parameters for dark energy, the total cosmic fluid, and the $\Lambda$CDM total effective fluid, for the case $\alpha=1$. Upper-right panel: same quantities, but for $\alpha=-1$. Lower-left panel: Hubble expansion rate $H(z)$ compared with Cosmic Chronometer data. The theoretical predictions for $\alpha=1$ and $\alpha=-1$ are both included, although they are practically indistinguishable at the plotted scale and therefore appear superposed. Lower-right panel: distance modulus $\mu(z)$ compared with the PantheonPlus + Cepheid supernova sample. As in the previous panel, the curves for $\alpha=1$ and $\alpha=-1$ are both shown, but they overlap almost completely and are visually superposed.}
\label{fig:cuatroimagenes}
\end{figure}

\subsection{Numerical results}

In this subsection, we perform a qualitative analysis of the model by considering the representative choices $\alpha=1$ and $\alpha=-1$, while fixing the remaining parameters to $\sigma=0.1$, $\lambda=0.1$, and $n=3$. The purpose of this study is to explore the dynamical behavior of the system at the background level before proceeding to a full statistical analysis. In contrast to the following section, where the model parameters will be constrained through a $\chi^2$ minimization using observational data, here we restrict our attention to particular parameter choices. This allows us to examine the structure of the phase space and the evolution of the relevant cosmological quantities, such as the fractional energy densities and the effective equation-of-state parameters.

This preliminary analysis provides insight into the viability of the model by assessing whether it can reproduce a consistent cosmological history, including the expected sequence of radiation-, matter, and dark energy-dominated epochs, as well as a late-time accelerated expansion. It also serves to identify potential pathological behaviors or inconsistencies that could arise for specific values of $\alpha$, thereby guiding the subsequent parameter estimation.

Fig.(\ref{Fig1-1}) shows the evolution curves in the phase space for the non-interacting model, for two different values of the parameter $\alpha$ (positive and negative) associated with the standard kinetic term, assuming that the power $n$ of the non-standard kinetic term is fixed to $n=3$. The left panel shows the evolution curves in the phase space for the case $\alpha=+1$ with the parameters $\lambda=0.1$ and $\sigma=0.1$, and the initial conditions are $x_i= 10 \times 10^{-6}$, $y_i=8.54 \times 10^{-13}$, $u_i=5 \times 10^{-5} $ and $\varrho_i=0.99983$. From the left panel in which $\alpha=+1$, we note that phase space stream flow from the trajectories $a_R\to e_M$ towards the attractor $h$ describes the evolution of the system. By considering the stability analysis of the critical points in the autonomous system, we observe that the system evolves toward the attractor $h$.
Analogously, in the right panel where the parameter $\alpha=-1$ with the parameters $\lambda=0.1$ and $\sigma=0.1$ and the initial conditions are $x_i= 10 \times 10^{-12}$, $y_i=5.1 \times 10^{-15}$, $u_i=1.179 \times 10^{-12} $ and $\varrho_i=0.99983$, we show that the phase space stream flow illustrates the trajectories $a_R\to e_M\to j^+$.  In this case, assuming the stability analysis of the critical points in the autonomous system, we note that, in particular, the system evolves toward the attractors $j^\pm$ (see Eq.(\ref{jj})).
In both plots ($\alpha=+1$ and $\alpha=-1$), the stability conditions corroborate that these critical points correspond to stable solutions associated with a dark energy-dominated epoch. Also, these plots show that in the case of the non-interacting model, it evolves toward a late-time attractor scenario, thereby ensuring a consistent description  of cosmic accelerated expansion.

In addition, for the non-interacting model,  in Fig.~\ref{fig:cuatroimagenes} the upper-left panel displays the evolution of the fractional energy densities of dark energy, matter, and radiation, together with the equation-of-state parameters for dark energy, the total cosmic fluid, and the $\Lambda$CDM total effective fluid, for the case $\alpha=1$. In this case, the model successfully reproduces the standard cosmological sequence, with a radiation-dominated era at early times, followed by a matter-dominated epoch, and finally a late-time acceleration driven by dark energy. The EoS parameter of dark energy approaches $w_{\rm de}\simeq -1$ at late times, mimicking a cosmological constant, while the total equation of state transitions consistently toward negative values, ensuring accelerated expansion. The upper-right panel shows the same quantities for $\alpha=-1$. Although the background evolution still exhibits the expected sequence of cosmological epochs, noticeable differences arise in the behavior of the dark energy equation of state. In particular, while $w_{\rm de}$ approaches values close to $-1$ at the present epoch, it deviates significantly at earlier times, entering a phantom-like regime.

The lower-left panel presents the Hubble expansion rate $H(z)$ compared with Cosmic Chronometer data. The theoretical predictions for $\alpha=1$ and $\alpha=-1$ are both included; however, they are practically indistinguishable at the plotted scale and therefore appear superposed, indicating that both parameter choices provide an equally good description of the expansion history at the background level. Similarly, the lower-right panel shows the distance modulus $\mu(z)$ compared with the PantheonPlus + Cepheid supernova sample. As in the previous panel, the curves corresponding to $\alpha=1$ and $\alpha=-1$ overlap almost completely and are visually superposed, reinforcing the degeneracy between these scenarios when only background observables are considered.

\section{Statistical analysis: $\chi^2$ estimators}
\label{sec8}

We now confront the different models with late-time cosmological data in order to constrain their parameter space. The analysis relies on three complementary probes: Cosmic Chronometers (CC), Type Ia Supernovae (SNe Ia), and Baryon Acoustic Oscillations (BAO) from the DESI DR2 release. Because each probe carries distinct information about the cosmic expansion, their joint use enables a self-consistent reconstruction of the expansion history of the Universe.
 
\subsection{General statistical framework}
 
For every dataset, we build the likelihood from a chi-square estimator of the form
\begin{equation}
\chi^2 = (\mathbf{D}_{\rm obs} - \mathbf{D}_{\rm th})^{T}
\,\mathbf{C}^{-1}\,
(\mathbf{D}_{\rm obs} - \mathbf{D}_{\rm th}),
\end{equation}
where $\mathbf{D}_{\rm obs}$ is the measured data vector and $\mathbf{D}_{\rm th}$ its theoretical prediction, evaluated for a given choice of model parameters $\mathbf{p}$, while $\mathbf{C}$ is the associated covariance matrix. When the measurements are mutually uncorrelated, this expression simplifies to a sum over the individual data points. The specific form of $\mathbf{D}_{\rm obs}$ and $\mathbf{D}_{\rm th}$ for each probe is detailed below.
 
\subsubsection{Cosmic Chronometers}
 
CC provides model-independent measurements of the Hubble parameter from the differential aging of passively evolving galaxies. They exploit the relation $H(z) = -\frac{1}{1+z}\frac{dz}{dt}$, which gives the expansion rate in terms of the redshift $z$ without relying on any integrated distance.
 
We use a compilation of $N_{\rm CC}$ data points $\{z_i, H_{\rm obs}(z_i), \sigma_{H_i}\}$ drawn from the literature \cite{Moresco:2020fbm,cao2018cosmological,farooq2013hubble}. Since the observable is $H(z)$ itself, the corresponding chi-square takes the simple form
\begin{equation}
\chi^2_{\rm CC} =
\sum_{i=1}^{N_{\rm CC}}
\frac{\left[H_{\rm th}(z_i;\mathbf{p}) - H_{\rm obs}(z_i)\right]^2}{\sigma_{H_i}^2}.
\end{equation}

\subsubsection{Type Ia Supernovae}
 
Type Ia supernovae probe the luminosity distance through the distance modulus. The observed modulus is defined as
\begin{equation}
\mu_{\rm obs} = m_B^{\rm corr} - M,
\end{equation}
whereas its theoretical value reads
\begin{equation}
\mu_{\rm th}(z;\mathbf{p}) = 5 \log_{10}\!\left[\frac{D_L(z;\mathbf{p})}{\mathrm{Mpc}}\right] + 25,
\end{equation}
with the luminosity distance given by
\begin{equation}
D_L(z;\mathbf{p}) = (1+z)\int_0^z \frac{c\,dz'}{H(z';\mathbf{p})}.
\end{equation}
 
We adopt the PantheonPlus + SH0ES compilation \cite{Brownsberger:2021uue,Brout:2022vxf,Scolnic:2021amr}\footnote{Available at \url{https://github.com/PantheonPlusSH0ES}.}, keeping the 1623 supernovae with $z>0.01$. The chi-square then reads
\begin{equation}
\chi^2_{\rm SN} =
(\boldsymbol{\mu}_{\rm obs} - \boldsymbol{\mu}_{\rm th})^{T}
\,\mathbf{C}^{-1}\,
(\boldsymbol{\mu}_{\rm obs} - \boldsymbol{\mu}_{\rm th}),
\end{equation}
with $\mathbf{C}$ accounting for both statistical and systematic uncertainties.
 
Because supernovae are sensitive only to relative distances, they cannot fix the absolute scale of the cosmic expansion. This leaves a degeneracy between $M$ and $H_0$, which is broken only once SNe Ia are combined with probes that either measure $H(z)$ directly or anchor absolute distances.

\subsubsection{Baryon Acoustic Oscillations}
 
BAO measurements act as a geometric ruler, comparing the observed clustering scale with the sound horizon at the drag epoch. We employ the DESI DR2 dataset \cite{DESI:2025zgx,DESI:2025zpo}\footnote{Available at \url{https://github.com/CobayaSampler/bao_data/}.}, which delivers both transverse and radial distance information.
 
The data vector reads
\begin{equation}
\mathbf{X}_{\rm obs} =
\left\{
\frac{D_M(z)}{r_d},\,
\frac{D_H(z)}{r_d},\,
\frac{D_V(z)}{r_d}
\right\}_{\rm obs},
\end{equation}
with the relevant distances given by
\begin{align}
D_M(z) &= \int_0^{z} \frac{c}{H(z')}\,dz', \\
D_H(z) &= \frac{c}{H(z)}, \\
D_V(z) &= \left[ D_M^2(z)\,\frac{c\,z}{H(z)} \right]^{1/3}.
\end{align}
 
The chi-square then follows as
\begin{equation}
\chi^2_{\rm DESI} =
(\mathbf{X}_{\rm obs} - \mathbf{X}_{\rm th})^{T}
\,\mathbf{C}^{-1}\,
(\mathbf{X}_{\rm obs} - \mathbf{X}_{\rm th}),
\end{equation}
where $\mathbf{C}$ is the covariance matrix released by the DESI collaboration. Throughout, the drag-epoch sound horizon is fixed to its $\Lambda$CDM value, $r_d = 147.09\,\mathrm{Mpc}$, in agreement with the Planck results \cite{Planck:2018vyg}.

\begin{figure}[bp]
\centering
\includegraphics[width=0.45\textwidth]{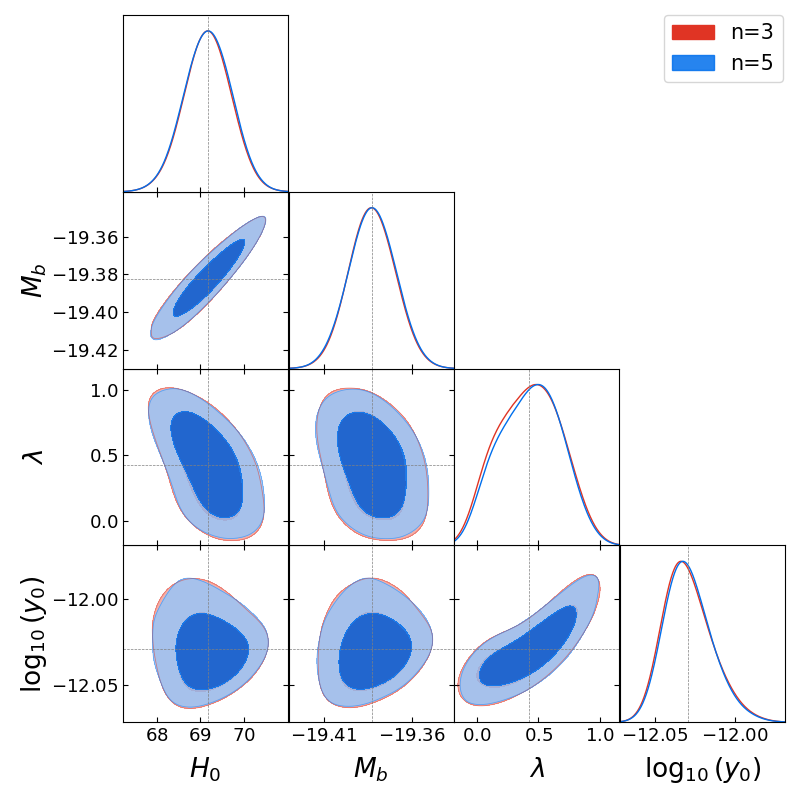}
\hfill
\includegraphics[width=0.45\textwidth]{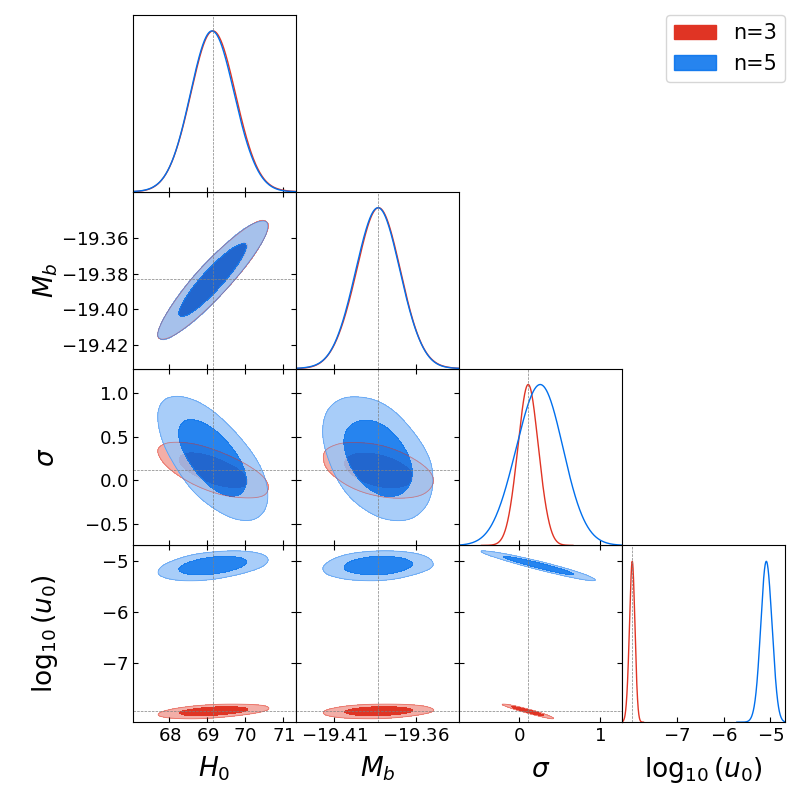}
\caption{\footnotesize
One- and two-dimensional marginalized posterior distributions at the 68\% and 95\% confidence levels obtained from the joint analysis of CC, PantheonPlus supernovae, and DESI DR2 BAO data for the non-interacting model. Left: Results for the $\alpha=1$ branch. Right: Results for the $\alpha=-1$ branch. In both panels, the red contours correspond to $n=3$, while the blue contours represent $n=5$.
}
\label{fig:triangle_n3_n5_4param_desiDR2}
\end{figure}

\begin{table*}[tp]
\centering
\caption{Constraints from CC + PantheonPlus + SH0ES + DESI DR2 for the non-interacting model with $\alpha=\pm 1$, where two different values of the parameter $n$ are considered for each value of $\alpha$. We report mean values with $68\%$ confidence levels, best-fit values for the free parameters, the corresponding best-fit derived value of $\Omega_{m0}$, and goodness-of-fit statistics.}
\label{tab:combined_alpha_constraints}

\renewcommand{\arraystretch}{1.1}

\begin{tabular}{|c|c|c|c|c|}
\hline\hline

& \multicolumn{2}{c|}{$\alpha = 1$}
& \multicolumn{2}{c|}{$\alpha = -1$} \\
\cline{2-5}

\textbf{Parameter}
& $n = 3$
& $n = 5$
& $n = 3$
& $n = 5$ \\

\hline

\multicolumn{5}{|c|}{\textit{Mean values and $68\%$ CL}} \\
\hline

$H_0$ 
& $69.17^{+0.49}_{-0.48}$ 
& $69.18^{+0.49}_{-0.49}$ 
& $69.16^{+0.53}_{-0.53}$ 
& $69.14^{+0.53}_{-0.53}$ \\

$M_b$ 
& $-19.383^{+0.012}_{-0.012}$ 
& $-19.382^{+0.012}_{-0.012}$ 
& $-19.383^{+0.012}_{-0.012}$ 
& $-19.383^{+0.012}_{-0.012}$ \\

$\lambda$ 
& $0.427^{+0.274}_{-0.269}$ 
& $0.435^{+0.272}_{-0.254}$ 
& --- 
& --- \\

$\sigma$ 
& --- 
& --- 
& $0.112^{+0.118}_{-0.118}$ 
& $0.251^{+0.258}_{-0.263}$ \\

$\log_{10}y_0$
& $-12.029^{+0.012}_{-0.018}$ 
& $-12.029^{+0.012}_{-0.017}$ 
& --- 
& --- \\

$\log_{10}u_0$
& --- 
& --- 
& $-7.948^{+0.053}_{-0.053}$ 
& $-5.088^{+0.109}_{-0.106}$ \\

\hline

\multicolumn{5}{|c|}{\textit{Best-fit values}} \\
\hline

$H_0$ 
& $69.144$ 
& $69.140$ 
& $69.152$ 
& $69.212$ \\

$M_b$ 
& $-19.383$ 
& $-19.383$ 
& $-19.383$ 
& $-19.382$ \\

$\lambda$ 
& $0.516$ 
& $0.521$ 
& --- 
& --- \\

$\sigma$
& --- 
& --- 
& $0.114$ 
& $0.222$ \\

$\log_{10}y_0$
& $-12.027$ 
& $-12.027$ 
& --- 
& --- \\

$\log_{10}u_0$
& --- 
& --- 
& $-7.948$ 
& $-5.075$ \\

\hline

\multicolumn{5}{|c|}{\textit{Derived quantity (best fit)}} \\
\hline

$\Omega_{m0}$ 
& $0.291$ 
& $0.291$ 
& $0.342$ 
& $0.354$ \\

\hline

\multicolumn{5}{|c|}{\textit{Goodness of fit}} \\
\hline

$\chi^2_{\min}$ 
& $1473.68$ 
& $1473.68$ 
& $1473.75$ 
& $1473.77$ \\

$\chi^2_{\nu}$ 
& $0.886$ 
& $0.886$ 
& $0.886$ 
& $0.886$ \\

\hline\hline
\end{tabular}
\end{table*}

\begin{figure}[tp]
\centering
\includegraphics[width=0.49\textwidth]{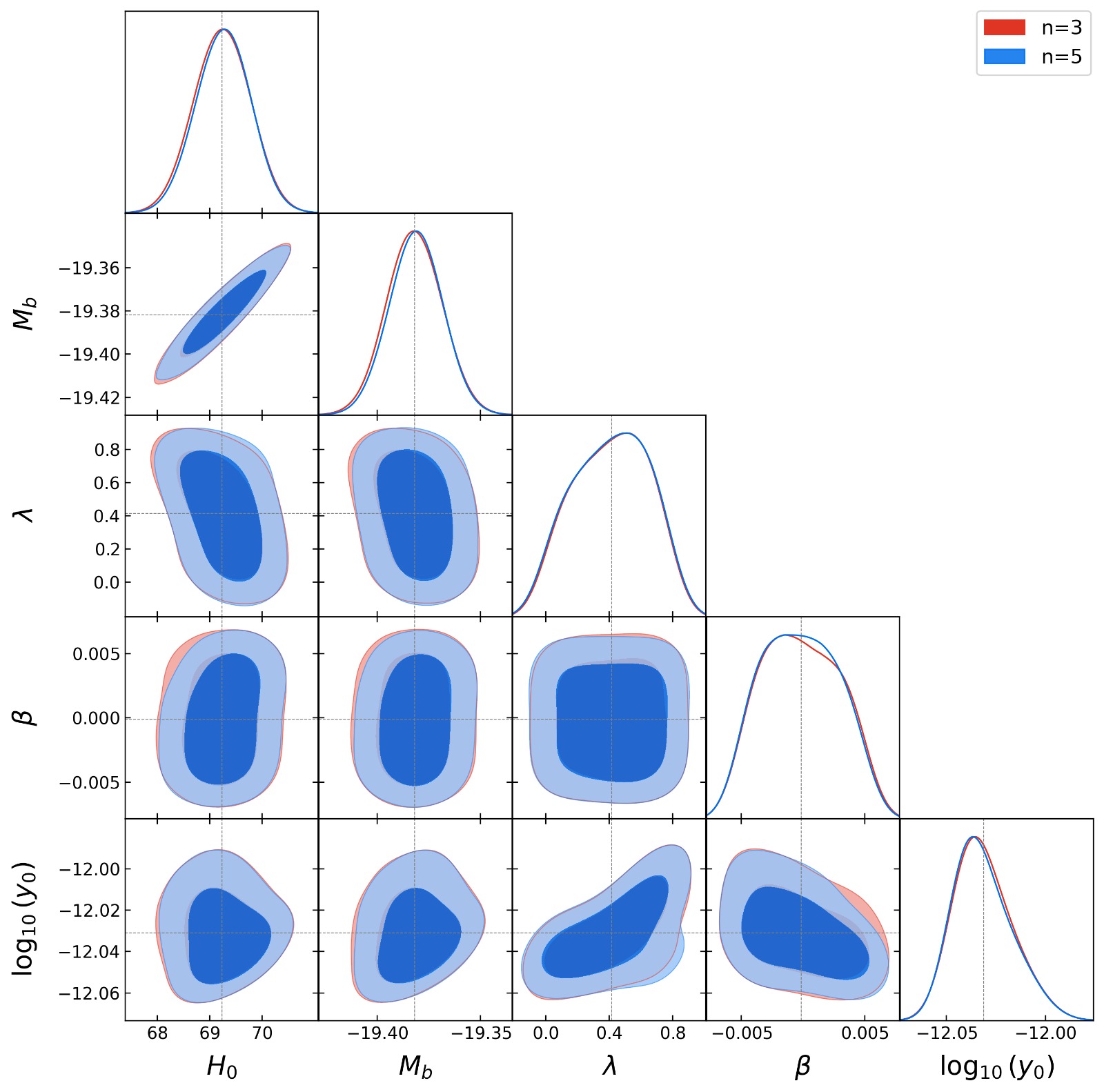}
\hfill
\includegraphics[width=0.49\textwidth]{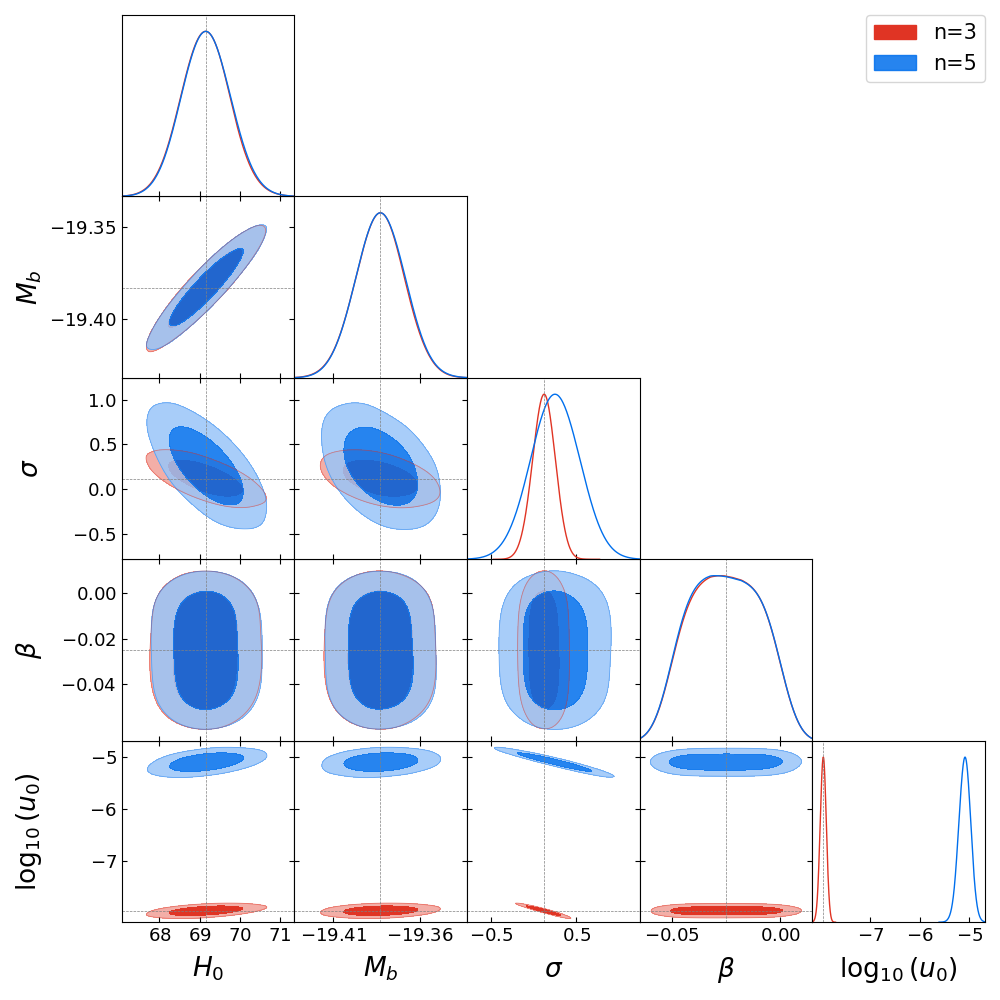}
\caption{\footnotesize
One- and two-dimensional marginalized posterior distributions at the 68\% and 95\% confidence levels obtained from the joint analysis of CC, PantheonPlus supernovae, and DESI DR2 BAO data for interaction $Q =\beta  \kappa \rho_m \dot\phi$. Left: Results for the $\alpha=1$ branch. Right: Results for the $\alpha=-1$ branch. In both panels, the red contours correspond to $n=3$, while the blue contours represent $n=5$.
}
\label{fig:triangle_n3_n5_5param_inter1_desiDR2}
\end{figure}

\begin{table*}[tp]
\centering
\caption{Constraints from CC + PantheonPlus + SH0ES + DESI DR2 for the interacting model 
$Q = \beta \kappa \rho_m \dot\phi$, considering different values of $n$ and 
$\alpha=\pm1$. We report mean values with $68\%$ confidence levels, best-fit values 
for the free parameters, the corresponding best-fit derived value of $\Omega_{m0}$, 
and goodness-of-fit statistics. }
\label{tab:combined_alpha_constraints_int1}

\renewcommand{\arraystretch}{1.1}

\begin{tabular}{|c|c|c|c|c|}
\hline\hline

& \multicolumn{2}{c|}{$\alpha = 1$}
& \multicolumn{2}{c|}{$\alpha = -1$} \\
\cline{2-5}

\textbf{Parameter}
& $n = 3$
& $n = 5$
& $n = 3$
& $n = 5$ \\

\hline

\multicolumn{5}{|c|}{\textit{Mean values and $68\%$ CL}} \\
\hline

$H_0$ 
& $69.24^{+0.48}_{-0.49}$ 
& $69.27^{+0.47}_{-0.47}$ 
& $69.15^{+0.53}_{-0.55}$ 
& $69.16^{+0.55}_{-0.55}$ \\

$M_b$ 
& $-19.382^{+0.012}_{-0.012}$ 
& $-19.381^{+0.012}_{-0.012}$ 
& $-19.383^{+0.013}_{-0.013}$ 
& $-19.383^{+0.013}_{-0.013}$ \\

$\beta$
& $-0.0003^{+0.003}_{-0.004}$ 
& $-0.0003^{+0.003}_{-0.004}$ 
& $-0.025^{+0.017}_{-0.017}$ 
& $-0.025^{+0.017}_{-0.017}$ \\

$\lambda$ 
& $0.415^{+0.280}_{-0.236}$ 
& $0.413^{+0.285}_{-0.239}$ 
& --- 
& --- \\

$\sigma$ 
& --- 
& --- 
& $0.120^{+0.120}_{-0.120}$ 
& $0.251^{+0.261}_{-0.257}$ \\

$\log_{10}y_0$
& $-12.031^{+0.012}_{-0.017}$ 
& $-12.031^{+0.012}_{-0.018}$ 
& --- 
& --- \\

$\log_{10}u_0$
& --- 
& --- 
& $-7.952^{+0.055}_{-0.055}$ 
& $-5.090^{+0.106}_{-0.108}$ \\

\hline

\multicolumn{5}{|c|}{\textit{Best-fit values}} \\
\hline

$H_0$ 
& $69.098$ 
& $69.151$ 
& $69.225$ 
& $69.192$ \\

$M_b$ 
& $-19.383$ 
& $-19.383$ 
& $-19.382$ 
& $-19.382$ \\

$\beta$
& $0.004$ 
& $-0.002$ 
& $-0.022$ 
& $-0.022$ \\

$\lambda$ 
& $0.557$ 
& $0.510$ 
& --- 
& --- \\

$\sigma$
& --- 
& --- 
& $0.103$ 
& $0.217$ \\

$\log_{10}y_0$
& $-12.029$ 
& $-12.025$ 
& --- 
& --- \\

$\log_{10}u_0$
& --- 
& --- 
& $-7.945$ 
& $-5.074$ \\

\hline

\multicolumn{5}{|c|}{\textit{Derived quantity (best fit)}} \\
\hline

$\Omega_{m0}$ 
& $0.279$ 
& $0.291$ 
& $0.327$ 
& $0.339$ \\

\hline

\multicolumn{5}{|c|}{\textit{Goodness of fit}} \\
\hline

$\chi^2_{\min}$ 
& $1473.66$ 
& $1473.65$ 
& $1473.76$ 
& $1473.75$ \\

$\chi^2_{\nu}$ 
& $0.886$ 
& $0.887$ 
& $0.886$ 
& $0.886$ \\

\hline\hline
\end{tabular}
\end{table*}

\begin{figure}[tp]
\centering
\includegraphics[width=0.49\textwidth]{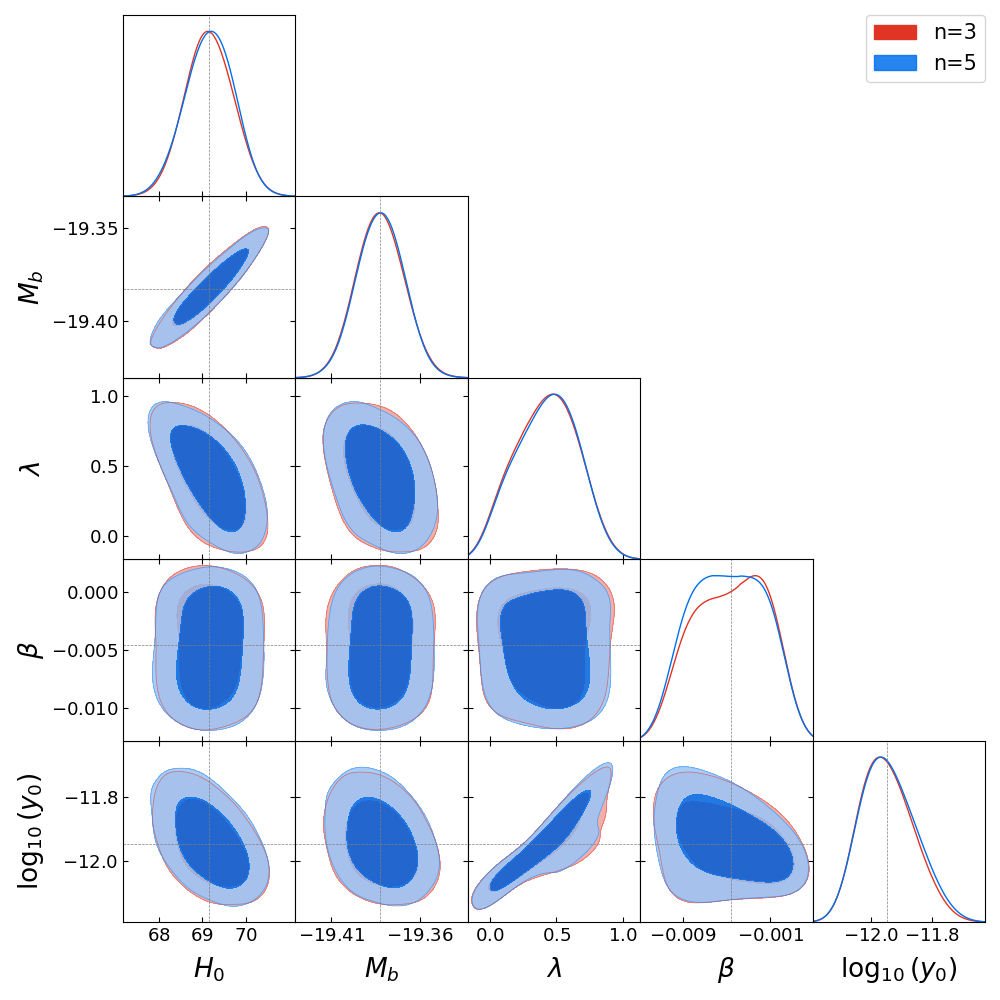}
\hfill
\includegraphics[width=0.49\textwidth]{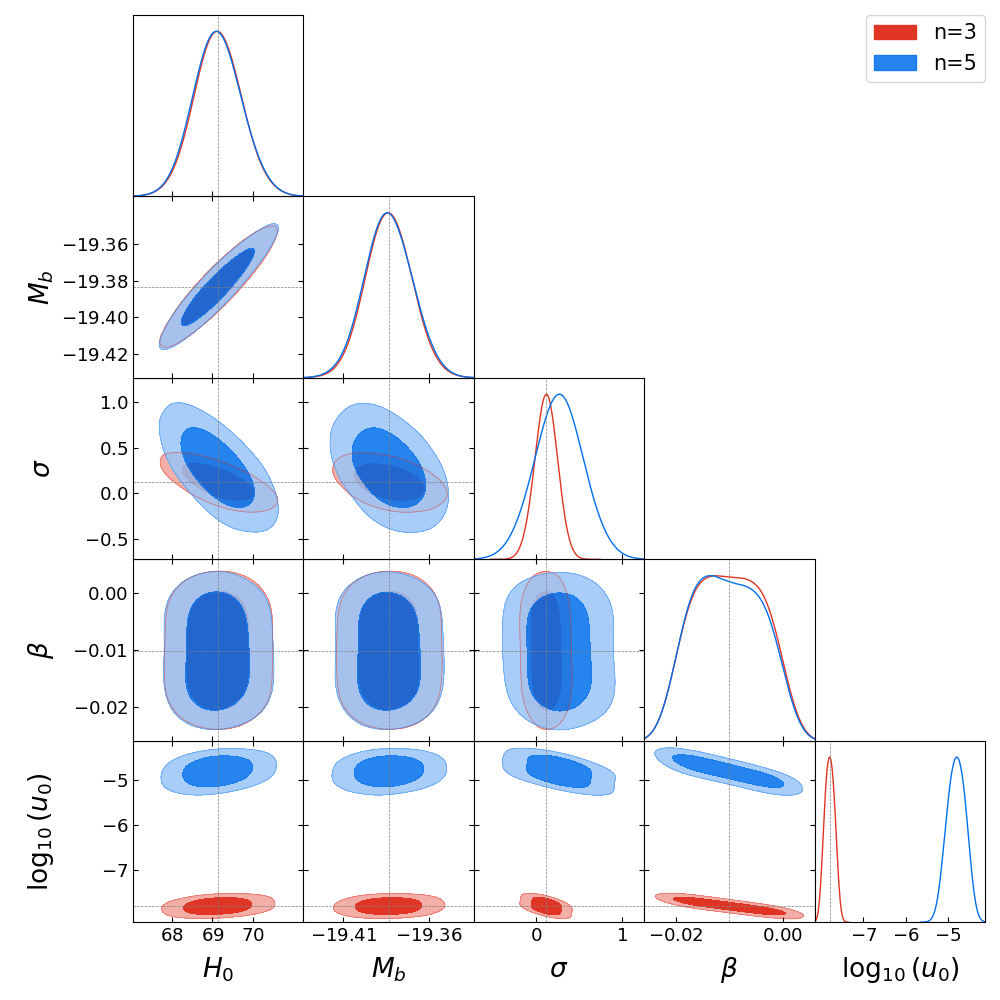}
\caption{\footnotesize
One- and two-dimensional marginalized posterior distributions at the 68\% and 95\% confidence levels obtained from the joint analysis of CC, PantheonPlus supernovae, and DESI DR2 BAO data for interaction  $Q \propto \rho_m H$. Left: Results for the $\alpha=1$ branch. Right: Results for the $\alpha=-1$ branch. In both panels, the red contours correspond to $n=3$, while the blue contours represent $n=5$.
}
\label{fig:triangle_n3_n5_5param_inter2_desiDR2}
\end{figure}

\begin{table*}[tp]
\centering
\caption{Constraints from CC + PantheonPlus + SH0ES + DESI DR2 for the interacting model 
$Q = 3 \beta \rho_m H$, considering different values of $n$ and 
$\alpha=\pm1$. We report mean values with $68\%$ confidence levels, best-fit values 
for the free parameters, the corresponding best-fit derived value of $\Omega_{m0}$, 
and goodness-of-fit statistics. }
\label{tab:combined_alpha_constraints_int2}

\renewcommand{\arraystretch}{1.1}

\begin{tabular}{|c|c|c|c|c|}
\hline\hline

& \multicolumn{2}{c|}{$\alpha = 1$}
& \multicolumn{2}{c|}{$\alpha = -1$} \\
\cline{2-5}

\textbf{Parameter}
& $n = 3$
& $n = 5$
& $n = 3$
& $n = 5$ \\

\hline

\multicolumn{5}{|c|}{\textit{Mean values and $68\%$ CL}} \\
\hline

$H_0$ 
& $69.16^{+0.51}_{-0.50}$ 
& $69.17^{+0.59}_{-0.54}$ 
& $69.14^{+0.53}_{-0.52}$ 
& $69.13^{+0.58}_{-0.62}$ \\

$M_b$ 
& $-19.383^{+0.012}_{-0.012}$ 
& $-19.382^{+0.012}_{-0.012}$ 
& $-19.383^{+0.012}_{-0.012}$ 
& $-19.384^{+0.013}_{-0.013}$ \\

$\beta$
& $-0.005^{+0.004}_{-0.003}$ 
& $-0.005^{+0.004}_{-0.003}$ 
& $-0.010^{+0.007}_{-0.007}$ 
& $-0.010^{+0.007}_{-0.007}$ \\

$\lambda$ 
& $0.424^{+0.260}_{-0.235}$ 
& $0.428^{+0.263}_{-0.227}$ 
& --- 
& --- \\

$\sigma$ 
& --- 
& --- 
& $0.120^{+0.118}_{-0.119}$ 
& $0.270^{+0.257}_{-0.259}$ \\

$\log_{10}y_0$
& $-11.946^{+0.076}_{-0.100}$ 
& $-11.940^{+0.079}_{-0.105}$ 
& --- 
& --- \\

$\log_{10}u_0$
& --- 
& --- 
& $-7.794^{+0.115}_{-0.114}$ 
& $-4.803^{+0.209}_{-0.211}$ \\

\hline

\multicolumn{5}{|c|}{\textit{Best-fit values}} \\
\hline

$H_0$ 
& $69.181$ 
& $69.134$ 
& $69.143$ 
& $69.169$ \\

$M_b$ 
& $-19.382$ 
& $-19.383$ 
& $-19.384$ 
& $-19.383$ \\

$\beta$
& $-0.007$ 
& $-0.009$ 
& $-0.019$ 
& $-0.015$ \\

$\lambda$ 
& $0.457$ 
& $0.462$ 
& --- 
& --- \\

$\sigma$
& --- 
& --- 
& $0.137$ 
& $0.271$ \\

$\log_{10}y_0$
& $-11.914$ 
& $-11.906$ 
& --- 
& --- \\

$\log_{10}u_0$
& --- 
& --- 
& $-7.658$ 
& $-4.655$ \\

\hline

\multicolumn{5}{|c|}{\textit{Derived quantity (best fit)}} \\
\hline

$\Omega_{m0}$ 
& $0.280$ 
& $0.278$ 
& $0.271$ 
& $0.284$ \\

\hline

\multicolumn{5}{|c|}{\textit{Goodness of fit}} \\
\hline

$\chi^2_{\min}$ 
& $1473.66$ 
& $1473.64$ 
& $1473.64$ 
& $1473.64$ \\

$\chi^2_{\nu}$ 
& $0.886$ 
& $0.886$ 
& $0.886$ 
& $0.886$ \\

\hline\hline
\end{tabular}
\end{table*}

\subsection{Discussion of the results}

For the \textit{non-interacting model}, the joint analysis of CC + PantheonPlus + SH0ES + DESI DR2 is summarized in Table~\ref{tab:combined_alpha_constraints} and illustrated in Fig.~\ref{fig:triangle_n3_n5_4param_desiDR2}. The background quantities are recovered with remarkable stability across all four configurations: $H_0 \simeq 69.1$--$69.2\,\mathrm{km\,s^{-1}\,Mpc^{-1}}$ and $M_b \simeq -19.38$, both essentially independent of $n$ and of the sign of $\alpha$. The derived matter density settles at $\Omega_{m0} \simeq 0.29$ for $\alpha = 1$ and at a marginally higher $\Omega_{m0} \simeq 0.34$--$0.35$ for $\alpha = -1$, both values being compatible with the standard $\Lambda$CDM result.

The branches differ mainly in which model-specific parameter the data are able to constrain. For $\alpha = 1$ the slope $\lambda$ is well measured ($\lambda \simeq 0.43^{+0.27}_{-0.26}$) while no bound can be placed on $\sigma$, whereas for $\alpha = -1$ the situation is reversed: $\sigma$ is constrained ($\sigma \simeq 0.11$--$0.25$) and $\lambda$ remains entirely undetermined. This selective behavior has a clear origin in the underlying dynamical system, where the late-time attractor that shapes the expansion history is governed by only one of these parameters in each branch, as reflected in the existence and stability conditions of the critical points reported in Tables~\ref{table1} and~\ref{table2c}. The parameter that is absent from the relevant fixed-point structure leaves no imprint on the background evolution and is therefore invisible to the observables employed here.

A dependence on $n$ is confined to the $\alpha = -1$ branch and is driven by the initial condition $\log_{10} u_0$, whose posterior shifts and broadens substantially as $n$ increases---from $\log_{10} u_0 = -7.95^{+0.05}_{-0.05}$ at $n=3$ to $-5.09^{+0.11}_{-0.11}$ at $n=5$---with the two distributions being essentially disjoint (Fig.~\ref{fig:triangle_n3_n5_4param_desiDR2}, right panel); the parameter $\sigma$ follows a milder version of the same trend. In the $\alpha = 1$ branch, by contrast, the $n=3$ and $n=5$ posteriors overlap almost perfectly, so the data are effectively insensitive to $n$. The quality of the fit is excellent throughout, with $\chi^2_{\min} \approx 1473.7$ and a reduced chi-square $\chi^2_\nu = 0.886$, placing the non-interacting model on par with $\Lambda$CDM in reproducing the late-time expansion. The small excess of the $\alpha = -1$ branch ($\chi^2_{\min} = 1473.75$--$1473.77$ against $1473.68$ for $\alpha = 1$) is statistically negligible, so neither branch nor any value of $n$ is preferred.

We now switch on the coupling $Q = \beta\,\kappa\,\rho_m\,\dot\phi$, which introduces $\beta$ as an additional free parameter while preserving the stability of the background sector, as summarized in Table~\ref{tab:combined_alpha_constraints_int1} and Fig.~\ref{fig:triangle_n3_n5_5param_inter1_desiDR2}. The background quantities are again recovered with high stability across the four configurations---$H_0 \simeq 69.2\,\mathrm{km\,s^{-1}\,Mpc^{-1}}$, $M_b \simeq -19.38$, and $\Omega_{m0} \simeq 0.28$--$0.29$ for $\alpha = 1$ and $\Omega_{m0}\simeq 0.33$--$0.34$ for $\alpha = -1$---and the same parameter selectivity persists: $\lambda$ is constrained only in the $\alpha = 1$ branch ($\lambda \simeq 0.41^{+0.28}_{-0.24}$) and $\sigma$ only in the $\alpha = -1$ branch ($\sigma \simeq 0.12$--$0.25$), while the initial condition $\log_{10} u_0$ again separates sharply between $n=3$ and $n=5$ for $\alpha = -1$ (from $-7.95^{+0.06}_{-0.06}$ to $-5.09^{+0.11}_{-0.11}$; Fig.~\ref{fig:triangle_n3_n5_5param_inter1_desiDR2}, right panel). As in the non-interacting model, this structure follows from the fixed-point analysis of the corresponding dynamical system, reported in Tables~\ref{table3} and~\ref{table4}. The truly new feature is the coupling $\beta$, which the data bound only weakly and find compatible with a vanishing interaction: $\beta \simeq -0.0003^{+0.003}_{-0.004}$ for $\alpha = 1$ and $\beta \simeq -0.025^{+0.017}_{-0.017}$ for $\alpha = -1$, both consistent with zero within $\sim 1.5\sigma$.

The second interaction, $Q = 3 \beta\,\rho_m\,H$, yields a picture fully analogous to the previous one, as shown in Table~\ref{tab:combined_alpha_constraints_int2} and Fig.~\ref{fig:triangle_n3_n5_5param_inter2_desiDR2}. The background parameters are again stable across the four configurations ($H_0 \simeq 69.2\,\mathrm{km\,s^{-1}\,Mpc^{-1}}$, $M_b \simeq -19.38$), and here the derived matter density is uniform across both branches, $\Omega_{m0} \simeq 0.27$--$0.28$, in contrast to the slightly higher $\alpha = -1$ values found for the previous coupling. The same parameter selectivity holds---$\lambda$ is bounded only for $\alpha = 1$ ($\lambda \simeq 0.42^{+0.26}_{-0.24}$) and $\sigma$ only for $\alpha = -1$ ($\sigma \simeq 0.12$--$0.27$)---and the initial condition $\log_{10} u_0$ once more splits between $n=3$ and $n=5$ in the $\alpha = -1$ branch (from $-7.79^{+0.12}_{-0.11}$ to $-4.80^{+0.21}_{-0.21}$), as dictated by the fixed-point structure of the dynamical system reported in Tables~\ref{table5} and~\ref{table6}. The coupling is again small and compatible with zero, $\beta \simeq -0.005^{+0.004}_{-0.003}$ for $\alpha = 1$ and $\beta \simeq -0.010^{+0.007}_{-0.007}$ for $\alpha = -1$, retaining the same negative sign. The fit is essentially identical to the non-interacting and first interacting cases, with $\chi^2_{\min} = 1473.64$--$1473.66$ and $\chi^2_\nu = 0.886$ throughout, confirming the absence of any statistical preference for this coupling.

Interestingly, none of the interacting configurations favor a positive coupling: $\beta$ is clearly negative in the $\alpha = -1$ branch and, in the $\alpha = 1$ branch, although compatible with zero, its posterior leans toward negative rather than positive values. This mild preference for a negative coupling is in line with recent post-DESI DR2 analyses of interacting dark energy: Ref.~\cite{Figueruelo:2026eis}, for instance, reports that scenarios with energy transfer from dark matter to dark energy---corresponding to a negative coupling---are mildly favored over $\Lambda$CDM. Present background data thus constrain the coupling to be small while consistently recovering this negative sign; establishing it as significantly nonzero will likely require complementary, perturbation-level probes.

\section{Conclusions}\label{conclusion}

In this article, we have studied a generalized interacting dilatonic ghost condensate model as a candidate for dark energy. The model is described by a  Lagrangian that incorporates two dominant kinetic terms,  one linear and one of arbitrary integer power $n>2$, with an exponential coupling function, together with an exponential potential $V(\phi)$. By allowing $n$ to remain a free parameter rather than fixing it to $n=2$ case of the standard dilatonic ghost condensate, the framework significantly broadens the class of dark energy models that can be explored within this type of scalar field theory.

 In order to analyze the dynamics of our model, we have considered the construction of an autonomous dynamical system using a carefully chosen set of dimensionless variables. This allowed us to identify and classify the critical points of the system for three distinct scenarios:  the non-interacting case $Q=0$ and two interacting models, where the energy transfer between dark energy and dark matter is governed by $Q\propto \rho_m\dot{\phi}$ and $Q\propto \rho_m H$, respectively. In all three cases, we have found that the critical points cover the full sequence of cosmological stages expected from a viable dark energy model: radiation domination, matter domination, and a late-time dark-energy-dominated accelerated expansion.

In our analysis, we have obtained the eigenvalues of the stability matrix around each critical point and determined the conditions under which these points behave as saddle points, unstable nodes, or stable attractors. In particular, the critical points $h$ and $j^\pm$ are related to the dark-energy-dominated solutions in which the density parameter associated with dark energy $\Omega_{de}=1$. For the situation in which the parameter $\alpha>0$, the critical point  $h$ describes quintessence behavior, and for $\alpha$ negative, it corresponds to the phantom regime. For the points $j^\pm$ emerge as stable attractors when $\alpha<0$ and appropriate conditions on $\lambda$, $\sigma$, and $n$ are satisfied. Here, we have noted that the interaction term $Q$ does not modify the stability conditions of these critical points, indicating that $h$ and $j^\pm$ are robust features of the model and remain stable features.

The phase-space analysis confirmed that for the different parameter choices, the system evolves from trajectory $a_R\to e_M \to (h$ or $j^\pm$ ), associated with radiation, matter, and finally settling into a late-time dark energy attractor. We have found that  the phase-space structure is preserved for both signs related to the parameter $\alpha$
with the key distinction that the identity of the late-time attractor, critical point $h$ for $\alpha$ positive or points $j^{\pm}$ for $\alpha<0$, and in the corresponding  EoS behavior, which leads to either 
 quintessence or phantom regimes at late times, respectively.

In order to constrain the different parameters related to the model with observational data, we carried out a statistical analysis considering a combined dataset of CC, Type Ia Supernovae from the PantheonPlus sample, and Baryon Acoustic Oscillation measurements from the DESI DR2 survey. In this analysis, we have considered two values of the exponent $n$ ($n=3$ and $n=5$), and both signs of the coupling parameter $\alpha$ associated with the standard kinetic term   ($\alpha=+1$ and $\alpha=-1$).  

For the \textit{non-interacting model}, the combined CC + PantheonPlus + SH0ES + DESI DR2 dataset yields robust and tightly constrained parameter estimates, with both branches exhibiting a remarkable level of mutual consistency, as shown in Table~\ref{tab:combined_alpha_constraints} and Fig.~\ref{fig:triangle_n3_n5_4param_desiDR2}. In this sense, we have found that all cases provide an equally good fit to the data, with comparable values of the minimum chi-square statistic. These results indicate that the non-interacting model in which $Q=0$ describes the late-time expansion history at a level comparable to $\Lambda$CDM, with no statistically significant preference for either branch or any particular value of $n$.

In relation to the first interaction $Q\propto \dot{\phi}\,\rho_m$, we have found that this interaction has a negligible impact on the cosmological constraints in relation to the non-interacting case. The results summarized in Table~\ref{tab:combined_alpha_constraints_int1} and Fig.~\ref{fig:triangle_n3_n5_5param_inter1_desiDR2} show that the values of the inferred parameters and their correlations remain largely unchanged. Also, we have obtained that the parameter $\lambda$ associated to the effective potential, is constrained only in the
case where $\alpha=1$ (for $n=3$ and $n=5$), whereas the parameter $\sigma$ related to the coupling parameter to $X^n$ remains unconstrained. However, for the case where the parameter $\alpha=-1$, we have found the opposite situation;  the parameter $\sigma$ is constrained, while   $\lambda$ remains unconstrained. With respect to the direction of the energy  flow, we have obtained  that 
both possibilities, $Q>0$ and $Q<0$, are allowed in the 
 $\alpha=+1$  branch. However, in the case where the parameter $\alpha=-1$, we have found that the direction of the energy flow is always negative, since the coupling parameter $\beta<0$.

With regard to the second interaction $Q\propto H\rho_m$ we have found that the different constraints on the observational parameters are analogous to the previous one, as shown in Table~\ref{tab:combined_alpha_constraints_int2} and Fig.~\ref{fig:triangle_n3_n5_5param_inter2_desiDR2}. As in the previous interaction, the parameter $\lambda$ is constrained only for the branch $\alpha=+1$, whereas the parameter $\sigma$ remains unstrained. In the $\alpha=-1$ branch, the situation is reversed; $\sigma$  is constrained, whereas $\lambda$ remains unconstrained. In relation to the direction of the energy flow, we have found that for both values of  $\alpha=\pm1$, the interaction parameter $\beta$   is always negative, implying that dark matter transfers energy to the dark energy component, i.e., $Q<0$. This result suggests that within the non-canonical framework described by the Lagrangian in Eq.(\ref{exp2}), the interaction $Q\propto H\rho_m$ accounts for a transfer of energy from DM to DE, in agreement with the behavior reported in other models with the same interaction, see Refs.\cite{Kumar:2016zpg,Pan:2023mie,Yang:2025uyv,Dai:2026pvx,Figueruelo:2026eis}.

Finally, a comprehensive study of the evolution of matter density perturbations within the framework of the generalized interacting dilatonic ghost condensate model presented here will be the subject of future work. Such an investigation will allow us to assess  the growth of large-scale structures and place further constraints on the model parameter space considering current and future cosmological observations.

\begin{acknowledgments}

R.H. acknowledges financial support from FOVI No. 250062.
\end{acknowledgments}


\bibliography{bio}

\begin{appendix}

\end{appendix}

\end{document}